\documentclass[twoside,twocolumn,9pt]{article}
\usepackage{extsizes}
\usepackage[super,sort&compress,comma]{natbib} 
\usepackage[version=3]{mhchem}
\usepackage[left=1.5cm, right=1.5cm, top=1.785cm, bottom=2.0cm]{geometry}
\usepackage{balance}
\usepackage{mathrsfs}
\usepackage{mathptmx}
\usepackage{sectsty}
\usepackage{graphicx} 
\usepackage{lastpage}
\usepackage[format=plain,justification=justified,singlelinecheck=false,font={stretch=1.125,small,sf},labelfont=bf,labelsep=space]{caption}
\usepackage{float}
\usepackage{ulem}
\usepackage{fancyhdr}
\usepackage{fnpos}
\usepackage[english]{babel}
\addto{\captionsenglish}{%
  
}
\usepackage{array}
\usepackage{droidsans}
\usepackage{charter}
\usepackage[T1]{fontenc}
\usepackage[usenames,dvipsnames]{xcolor}
\usepackage{setspace}
\usepackage[compact]{titlesec}
\usepackage{hyperref}

\usepackage{epstopdf}

\definecolor{cream}{RGB}{222,217,201}

\begin{document}

\pagestyle{fancy}
\thispagestyle{plain}
\fancypagestyle{plain}{
\renewcommand{\headrulewidth}{0pt}
}

\makeFNbottom
\makeatletter
\renewcommand\LARGE{\@setfontsize\LARGE{15pt}{17}}
\renewcommand\Large{\@setfontsize\Large{12pt}{14}}
\renewcommand\large{\@setfontsize\large{10pt}{12}}
\renewcommand\footnotesize{\@setfontsize\footnotesize{7pt}{10}}
\makeatother

\renewcommand{\thefootnote}{\fnsymbol{footnote}}
\renewcommand\footnoterule{\vspace*{1pt}%
\color{cream}\hrule width 3.5in height 0.4pt \color{black}\vspace*{5pt}} 
\setcounter{secnumdepth}{5}

\makeatletter 
\renewcommand\@biblabel[1]{#1}            
\renewcommand\@makefntext[1]%
{\noindent\makebox[0pt][r]{\@thefnmark\,}#1}
\makeatother 
\renewcommand{\figurename}{\small{Fig.}~}
\sectionfont{\sffamily\Large}
\subsectionfont{\normalsize}
\subsubsectionfont{\bf}
\setstretch{1.125} 
\setlength{\skip\footins}{0.8cm}
\setlength{\footnotesep}{0.25cm}
\setlength{\jot}{10pt}
\titlespacing*{\section}{0pt}{4pt}{4pt}
\titlespacing*{\subsection}{0pt}{15pt}{1pt}

\fancyfoot{}
\fancyfoot[LO,RE]{\vspace{-7.1pt}\includegraphics[height=9pt]{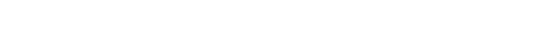}}
\fancyfoot[CO]{\vspace{-7.1pt}\hspace{13.2cm}\includegraphics{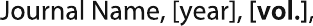}}
\fancyfoot[CE]{\vspace{-7.2pt}\hspace{-14.2cm}\includegraphics{head_foot/RF}}
\fancyfoot[RO]{\footnotesize{\sffamily{1--\pageref{LastPage} ~\textbar  \hspace{2pt}\thepage}}}
\fancyfoot[LE]{\footnotesize{\sffamily{\thepage~\textbar\hspace{3.45cm} 1--\pageref{LastPage}}}}
\fancyhead{}
\renewcommand{\headrulewidth}{0pt} 
\renewcommand{\footrulewidth}{0pt}
\setlength{\arrayrulewidth}{1pt}
\setlength{\columnsep}{6.5mm}
\setlength\bibsep{1pt}

\makeatletter 
\newlength{\figrulesep} 
\setlength{\figrulesep}{0.5\textfloatsep} 

\newcommand{\topfigrule}{\vspace*{-1pt}%
\noindent{\color{cream}\rule[-\figrulesep]{\columnwidth}{1.5pt}} }

\newcommand{\botfigrule}{\vspace*{-2pt}%
\noindent{\color{cream}\rule[\figrulesep]{\columnwidth}{1.5pt}} }

\newcommand{\dblfigrule}{\vspace*{-1pt}%
\noindent{\color{cream}\rule[-\figrulesep]{\textwidth}{1.5pt}} }

\makeatother

\twocolumn[
  \begin{@twocolumnfalse}
{\includegraphics[height=30pt]{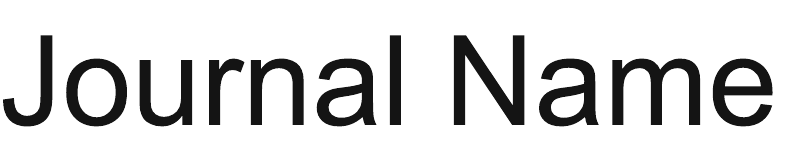}\hfill\raisebox{0pt}[0pt][0pt]{\includegraphics[height=55pt]{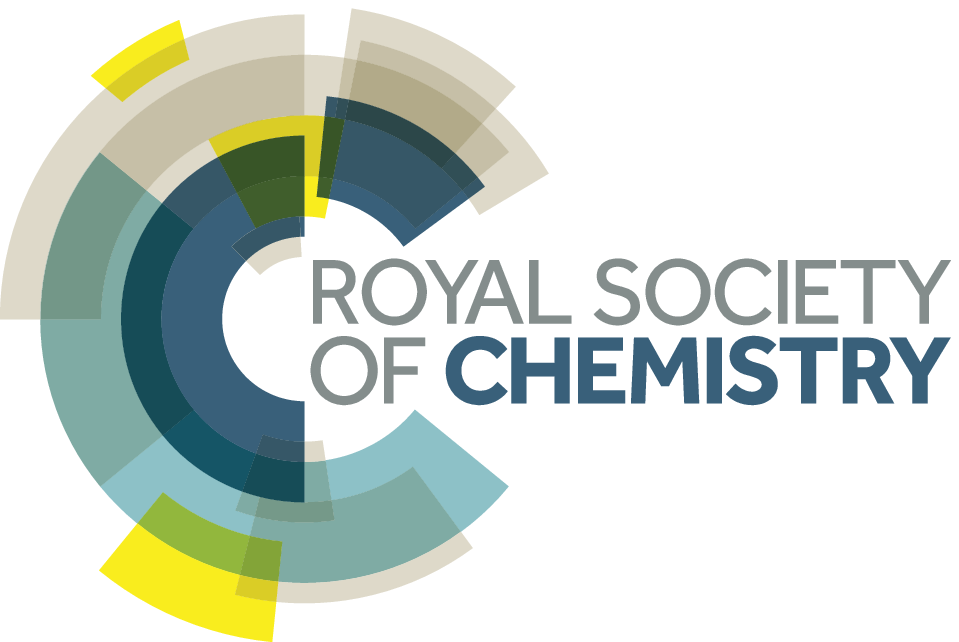}}\\[1ex]
\includegraphics[width=18.5cm]{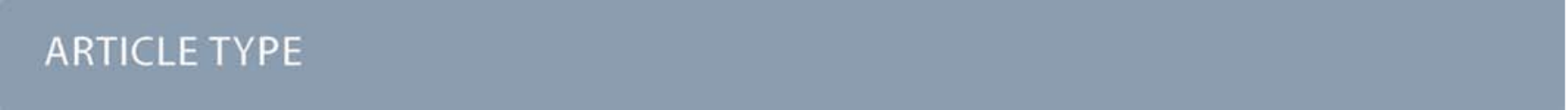}}\par
\vspace{1em}
\sffamily
\begin{tabular}{m{4.5cm} p{13.5cm} }

\includegraphics{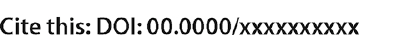} & \noindent\LARGE{\textbf{Ultraintense, ultrashort pulse x-ray scattering in small molecules$^\dag$}} \\
\vspace{0.3cm} & \vspace{0.3cm} \\

 & \noindent\large{Phay J. Ho,$^{\ast}$\textit{$^{a}$} Adam E. A. Fouda,$^{\ast}$\textit{$^{a}$}  Kai Li, \textit{$^{a,b}$} Gilles Doumy, \textit{$^{a}$} Linda Young,$^{\ast}$ \textit{$^{a,b,c}$}} \\

\includegraphics{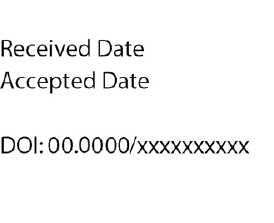} & \noindent\normalsize{We examine x-ray scattering from an isolated organic molecule from the linear to nonlinear absorptive regime. In the nonlinear regime, we explore the importance of both the elastic and inelastic channels and observe the onset of nonlinear behavior as a function of pulse duration and energy.  In the linear regime, we test the sensitivity of the scattering signal to molecular bonding and electronic correlation via calculations using the independent atom model (IAM), Hartree-Fock (HF) and density functional theory (DFT).  Finally, we describe how coherent x-ray scattering can be used to directly visualize femtosecond charge transfer and dissociation within a single molecule undergoing x-ray multiphoton absorption.}

\end{tabular}

 \end{@twocolumnfalse} \vspace{0.6cm}

  ]

\renewcommand*\rmdefault{bch}\normalfont\upshape
\rmfamily
\section*{}
\vspace{-1cm}


\footnotetext{\textit{$^{a}$~Chemical Sciences and Engineering Division, Argonne National Laboratory, Argonne, Illinois 60439, USA.}}
\footnotetext{\textit{$^{b}$~Department of Physics, The University of Chicago, Chicago, Illinois 60637, USA}}
\footnotetext{\textit{$^{c}$~James Franck Institute, The University of Chicago, Chicago, Illinois 60637, USA}}
\footnotetext{\textit{$^{*}$~ E-mail: pho@anl.gov, foudaae@anl.gov, young@anl.gov}}

\footnotetext{\dag~Electronic Supplementary Information (ESI) available: [Additional scattering calculations using alternative basis sets and molecular orientations.] See DOI: 00.0000/00000000.}



\section{Introduction}

Since the advent of x-ray free electron lasers that have conjured the dream of single molecule imaging \cite{Neutze-2000-Nature} via coherent x-ray scattering using “diffract-before-destroy” methodology, there has been considerable activity aimed at a quantitative understanding of x-ray-matter interactions in this novel high-fluence, x-ray multiphoton regime in free atoms \cite{Young-2010-Nature, Doumy-2011-PRL,Kanter-2011-PRL, Rudek-2012-NatPho, Fukuzawa-2013-PRL,Ho-2014-PRL}, molecules \cite{Rudenko-2017-Nature}, and small clusters \cite{Murphy-2014-NatComm}.  These studies have generally focused on interpreting ion and electron spectra resulting from the dominant photoabsorption channel, leaving the coherent (elastic) scattering channel less explored.   By contrast, coherent scattering has been the workhorse technique for studying larger targets, e.g. the biological community intent on optimizing femtosecond crystallography \cite{Chapman-2011-Nature} and single particle imaging (SPI) \cite{Seibert-2011-Nature}; and the atomic, molecular and optical physics community intent on understanding the dynamics of the nanoplasma environment associated with SPI using simple atomic and molecular cluster targets \cite{Gorkhover-2012-PRL,Ferguson-2016-ScienceAdv,Ho-2020-NatComm}.

Here we bridge this gap by studying computationally coherent x-ray scattering in small molecules from the linear to the x-ray multiphoton regime – a study inspired by the demonstration of a “molecular movie” of an electrocyclic ring-opening reaction via x-ray scattering from an ensemble of photoexcited gas-phase molecules \cite{Minitti-2015-PRL} on the $\sim100$-fs timescale.   Recent developments at XFELs of sub-femtosecond single-spike pulses in the hard \cite{Huang-2017-PRL,Marinelli-2017-APL} and the soft x-ray regime \cite{Duris-2020-NatPho} may allow one to probe processes on even shorter timescales, e.g. electron and wavepacket dynamics in small molecules, via coherent x-ray scattering \cite{Dixit-2012-PNAS}. Such scattering studies typically focus on following the electron and nuclear dynamics induced by photoexcitation from the valence shell.  However, photoexcitation of inner-shell electrons also induces ultrafast dynamics, a topic much less studied due to the challenge of x-ray pump --- x-ray probe experiments \cite{Picon-2016-NatComm,Liekhus-Schmaltz-2015-NatComm}.  We note that as an alternative to standard pump-probe experiments, it has recently been proposed to use the temporally stochastic $\sim1$-fs spikes within a single self-amplified spontaneous emission (SASE) pulse to sample multiple pump-probe delays and thus capture x-ray induced time-dependent dynamics using ghost imaging techniques \cite{Ratner-2019-PRX}.  Such an approach would be complementary to two-color x-ray pump-probe experiments which to date have been limited in time resolution to the few-femtosecond regime and have focused on x-ray processes in the linear regime. 

The nonlinear x-ray absorption regime is readily reached through focusing and understanding the associated x-ray scattering pattern is of foundational interest.  We investigate two situations for individual small molecules: (1) non-resonant, high-energy x-ray scattering from an organic molecule composed of only low-Z elements, and, (2) high-energy x-ray scattering from CH$_3$I where x-ray irradiation in the nonlinear absorptive regime has revealed a so-called "molecular black hole" due to rapid charging of the heavy atom and intramolecular electron transfer to the ionized site \cite{Rudenko-2017-Nature}.  For the first case, high energy x-ray scattering from a low-Z molecule, nonlinear absorption occurs only above the saturation fluence, which varies with photon energy.  There is an increasing importance of incoherent (Compton) scattering with x-ray energy, which unless accounted for, can swamp the desired coherent scattering signal \cite{Slowik-2014-NJP,Ho-2016-PRA}. For the second case, the high-Z atom acts as a sink for x-ray photons and we show that x-ray coherent scattering can provide a {\em{direct}} visualization of the intramolecular charge transfer and dissociation process.


In the nonlinear x-ray absorption regime, coherent x-ray scattering will include contributions from core-excited states and others produced within the pulse. In particular, the presence of a vacancy in a core-shell of a low-Z molecule can have a considerable effect on the molecular bonding and electron correlation, hence affecting the electron density and molecular structure of the molecule.  To investigate the feasibility of using coherent scattering to explore molecular bonding and electron correlation effects in a small molecule, we compare the coherent scattering patterns computed from the wave functions of these electronic states using different electronic structure theories: independent atom model (IAM), Hartree-Fock (HF) and density functional theory (DFT). In comparison to IAM, both HF and DFT provide a more accurate description of the electron density, in which HF includes the effect of molecular bonding, whereas DFT includes electron correlation \cite{lee1988development}.  

This contribution is organized as follows.  Section \ref{sec:method} describes our theoretical approaches to calculate coherent x-ray scattering from small molecules: Section \ref{ssec:MC/MD} describes our Monte Carlo/Molecular Dynamics code \cite{Ho-2017-JPB}, which implicitly uses the independent atom model (IAM) and which was previously used to calculate coherent scattering from sucrose molecules \cite{Ho-2020-NatComm, Ho-2016-PRA}. Section \ref{sec:adam} describes two \textit{ab initio} methods, Hartree-Fock (HF) and density functional theory (DFT), we use to describe the electron density within the molecule beyond the IAM. Section \ref{sec:Results} reports x-ray scattering from 1,3-cyclohexadiene. Section \ref{ssec:Results-MCMD} shows results using the MC/MD approach for coherent, incoherent and total scattering as a function of pulse fluence, pulse duration and photon energy (5.6, 9 and 24 keV). Section \ref{sec:compare} compares the coherent scattering patterns of 1,3-cyclohexadiene, in the low intensity limit, calculated with increasing accuracy: for the IAM-approximation, HF (addition of molecular bonding) and DFT (addition of electron correlation).  Section \ref{sec:CH3I} discusses how one may use linear x-ray scattering to probe x-ray multiphoton induced processes in small molecules - using CH$_3$I as it undergoes intramolecular charge transfer as an example.   Section \ref{sec:Outlook} presents a summary of our study and an outlook.     

\begin{figure*}[t]
\centering
  \includegraphics[width=5.5in] {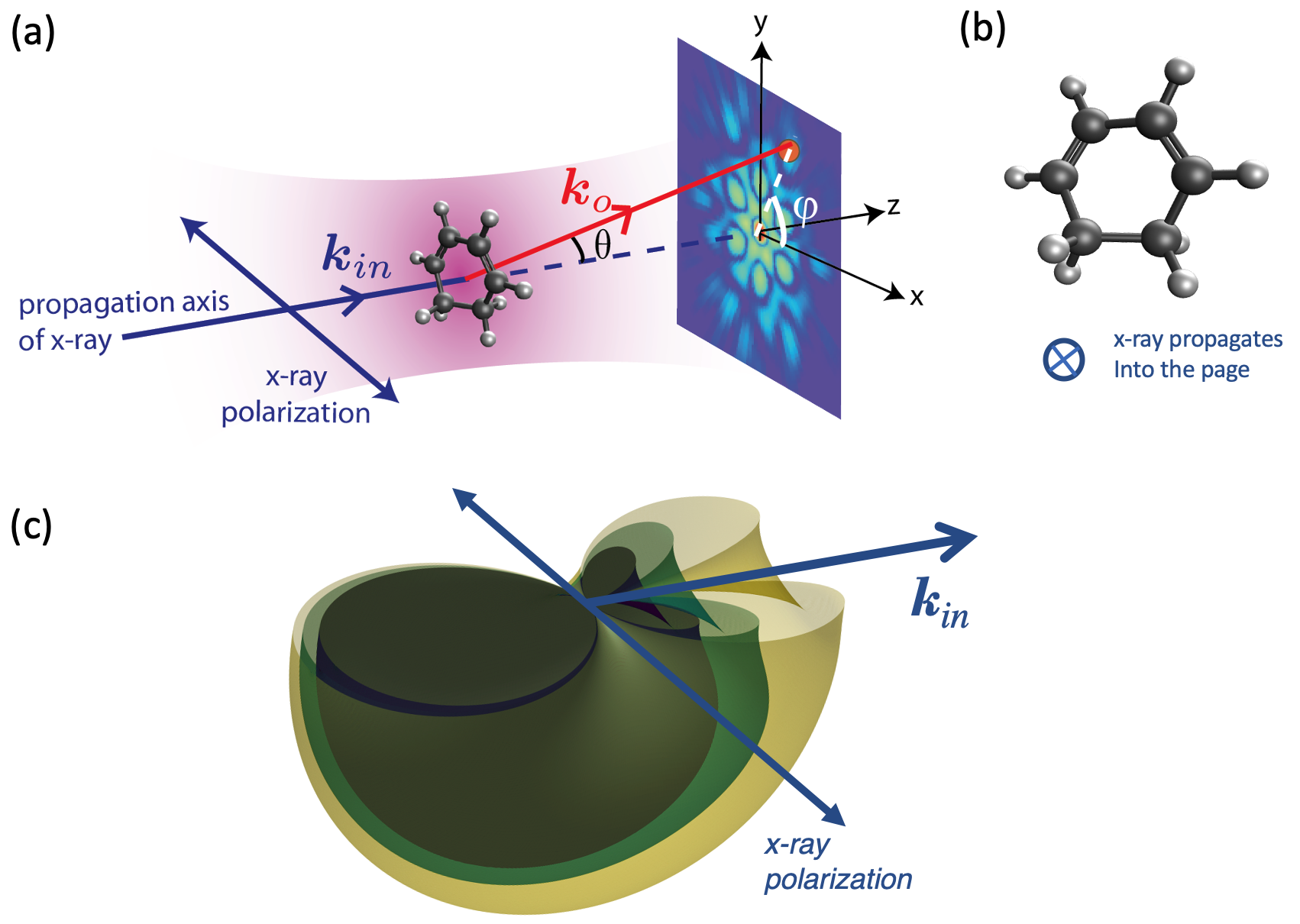}
 \caption{(a) Coherent x-ray scattering of 1,3-cyclohexadiane. (b) The orientation of the molecule with x rays propagating into the page.
 (c) Angular distribution of inelastic scattering of the molecule at 5.6 (purple curve), 9 (green curve) and 24 keV (yellow curve).}
  \label{fgr:Scattering}
\end{figure*}

\section{Theoretical Methods}\label{sec:method}


Since the seminal work of Hajdu and collaborators \cite{Neutze-2000-Nature}, many theoretical studies have been developed to investigate radiation damage processes of isolated biomolecules and clusters in intense XFEL pulses. Starting from sequential photoexcitation of inner-shell electrons at atomic sites, these pulses produce a multitude of transient electron states and induce complex correlated electron and nuclear dynamics in the sample volume on the timescale of an x-ray pulse.  All these dynamics can have a direct impact on the resolution limit of SPI.  In particular, many continuum models, which describe the target with atomic density, have been useful for studying the processes of radiation damage \cite{Hau-Riege-2004-PRE, Hau-Riege-2007-PRE,Ziaja-2006-EPJD,Moribayashi-2008-JPB,Quiney-2011-NatPhy, Caleman-2011-JMO} on large structures.  To provide an atomistic view and reveal the impact of sample size and heterogeneity on the induced x-ray response, a suite of complementary particle-based approaches \cite{Saalmann-2002-PRL, Jurek-2004-EPJD, Bergh-2004-PRE, Lorenz-2012-PRE, Gorobtsov-2015-PRE, Jurek-2016-JAC,Ho-2016-PRA, Yoon-2016-SR, Fortmann-Grote-2017-IUCRJ} has been developed.  Among these particle-based approaches, frozen nuclei approximation \cite{Lorenz-2012-PRE,Gorobtsov-2015-PRE} was used to examine the impact of electronic damage on both the elastic and inelastic channels in few-femtosecond XFEL pulses.  On the other hand, hybrid quantum/classical methods, like XMDYN \cite{Jurek-2016-JAC, Yoon-2016-SR, Fortmann-Grote-2017-IUCRJ} and MC/MD \cite{Ho-2016-PRA, Ho-2017-JPB}, are capable of tracking the correlated electron and nuclear dynamics and the changes in electronic configurations.  They have been used to simulate scattering patterns under realistic experimental conditions \cite{ Yoon-2016-SR, Fortmann-Grote-2017-IUCRJ} and capture the scattering response from electronic transients \cite{Ho-2020-NatComm}.

While these particle approaches have shown to provide insights into the effect of non-linear x-ray processes, they are mostly limited to independent atom model (IAM) for calculating the scattering response.  Recent work by Carrascosa and coworkers has examined the elastic and inelastic x-ray scattering of molecule in the weak-field limit \cite{Carrascosa-2019-JCTC}, in which it is shown that the scattering responses calculated from  \textit{ab initio} electronic wave functions can be very different from those predicted from the independent atom model.  Extending this approach to nonlinear x-ray regime is desired, but it remains a challenge as it requires tracking of a large number of full molecular electronic states, including both core-excited and valence-excited states.  Here, we compare the scattering responses of the ground state and core-excited states of 1,3-cyclohexadiene obtained using IAM with those computed from molecular wave functions derived with HF and DFT theories.

In this paper, we show results corresponding to a fixed molecular orientation in order to clearly illustrate the impact of non-linear x-ray interaction in the scattering images.  More importantly, this is because molecular structure information is largely washed out in the scattering image from an thermal ensemble of molecular gas due to rotational averaging \cite{Ho-2008-PRA, Ho-2009-JCP, Carrascosa-2019-JCTC}.  Even though perfect alignment without significant distortion of the electronic structure is not experimentally achievable at this point, a high degree of alignment can be obtained using an intense laser pulse and a cold molecular sample \cite{Stapelfeldt-2003-RMP,Spence-2004-PRL}. Our previous works showed that the scattering images from a ensemble of imperfectly aligned molecules achieved under both adiabatic and impulsive laser alignment conditions still retain the signatures of coherent scattering of a single molecule \cite{Ho-2008-PRA, Ho-2009-JCP,Ho-2009-JCP2}.  In addition, it is possible to reconstruct the scattering signal of fully aligned molecules from a series of measurements performed with imperfect alignment, so long as the degree of alignment is well characterized, as was demonstrated on ultrafast electron diffraction data from aligned molecules \cite{hensley2012imaging}.

\subsection{Independent Atom Model: MC/MD calculations} \label{ssec:MC/MD}


We model the molecular scattering response as a sum of the instantaneous scattering patterns weighted by the pulse intensity, $j_X(\tau,t)$ with FWHM duration $\tau$.  This model allows us to directly compare the response in both the linear and non-linear x-ray interaction regime, in which the incoming photons scatter from the initially prepared molecular state in the linear regime, whereas the incoming photons arriving at different times will scatter off the instantaneously populated transient states in the non-linear regime.  In our model, the scattering signals expressed in terms of the total differential cross section of the molecule can be regarded as the sum of the coherent (elastic) and incoherent (inelastic) scattering \cite{Hubbell-1975-JPCRD, Chihara-1987-JPF, Crowley-2014-HEDP2, Slowik-2014-NJP} 
\begin{eqnarray}
  \label{eq:totalSCS}
  \frac{d  \sigma_{total}}{d\Omega}(\boldsymbol{q}) =\frac{d  \sigma_{coh}}{d\Omega}(\boldsymbol{q}) \frac{d  \sigma_{comp}}{d\Omega}(\boldsymbol{q}) \ ,
\label{eq:totalDCS}
\end{eqnarray}
where coherent scattering can be expressed as
\begin{eqnarray}
  \label{eq:npSCS}
 \frac{d\sigma_{coh}}{d\Omega}(\boldsymbol{q}) &= & \frac{d\sigma_{KN}}{d\Omega} \frac{1}{\mathscr{F}}\int_{-\infty}^{+\infty}\!\! dt  j_X(\tau,t) |F^{(t)}_{mol}(\boldsymbol{q},t)|^2 \ ,
\end{eqnarray}
with $d\sigma_{KN}/d\Omega$ is the Klein-Nishina scattering cross section \cite{Klein-1929-ZFP} and $\mathscr{F} = \int_{-\infty}^{+\infty}\!\! dt  j_X(\tau, t)$ is the fluence of an XFEL pulse.  Here,
\begin{equation}
  \label{eq:timeFmol}
   F^{(t)}_{mol}(\boldsymbol{q},t) = \int d^{3} \boldsymbol{r} \rho_{1e}(\boldsymbol{r};\{\boldsymbol{R}_{j}\},t) e^{i\boldsymbol{q} \cdot\boldsymbol{r}}.
\end{equation}
is the time-dependent form factor of the the target molecule, where $\rho_{1e}(\boldsymbol{r};\{\boldsymbol{R}_{j}\},t)$ is the time-dependent electron density of the molecule with a geometry, $\{\boldsymbol{R}_{j}\}$, defined in the reference frame of the detector. In this reference frame, as shown in Fig. \ref{fgr:Scattering}, the momentum transfer vector $\boldsymbol{q}$ is expressed by,
\begin{equation}
  \label{eq:Q}
   \boldsymbol{q} = |\textbf{k}_{in}|\sin(\theta /2) \begin{pmatrix}-\cos(\phi)\cos(\theta/2)\\-\sin(\phi)\cos(\theta/2)\\\sin(\theta /2)\end{pmatrix},
\end{equation}
where,
\begin{equation}
  \label{eq:kin}
    \textbf{k}_{in} = \alpha\omega_x
\end{equation}
Here $\omega_x$ is the x-ray frequency and $\alpha$ is the fine structure constant.

By using the independent atom model, $F_{mol}(\boldsymbol{q},t)$ can be written as
\begin{equation}
  \label{eq:npFormFactor}
 F^{(t)}_{mol}(\boldsymbol{q},t)=\sum_{j=1}^{N_a} f_j(\boldsymbol{q},C_j(t))e^{i \boldsymbol{q} \cdot \boldsymbol{R}_j(t)} + \sum_{j=1}^{N_e(t)} e^{i \boldsymbol{q} \cdot \boldsymbol{r}_j(t)}\, ,
\end{equation}
where $N_a$ is the total number of atoms/ions, $\boldsymbol{R}_j(t)$, $C_j(t)$ and $f_j(\boldsymbol{q},C_j(t))$ are the position, the electronic configuration and the atomic form factor of the $j$-th atom/ion respectively.  $N_e(t)$ is the number of delocalized electrons within the focal region of the x-ray pulse, ($\boldsymbol{r}_j(t)$) are their positions.  For a long pulse, the energetically ejected electrons may escape beyond the focal area and will not contribute to the scattering signals. 

The contribution from incoherent scattering processes is cast in terms of the inelastic scattering function, $ S(\boldsymbol{q},t)$ \cite{Hubbell-1975-JPCRD}:
\begin{equation}
  \label{eq:comptonSCS}
  \frac{d  \sigma_{comp}}{d\Omega}(\boldsymbol{q}) =\frac{d\sigma_{KN}}{d\Omega}\frac{1}{\mathscr{F}}  \int_{-\infty}^{+\infty}\!\! dt  j_X(\tau,t) S(\boldsymbol{q},t) \, ,
\end{equation}
with 
\begin{equation}
  \label{eq:npisf}
  S(\boldsymbol{q},t)=\sum_{j=1}^{N_a} s_j(\boldsymbol{q},C_j(t)) \, ,
\end{equation}
and $s_j(\boldsymbol{q},C_j(t))$ is the inelastic scattering function of the $j$-th atom/ion with electronic configuration $C_j(t)$.  For our considered x-ray photon energies, which are much less the electron rest mass energy, $d\sigma_{KN}/d\Omega$ in eq. (\ref{eq:npSCS}) and (\ref{eq:comptonSCS}) can be approximated by Thomson differential scattering 
\begin{equation}
  \label{eq:thomson}
    \frac{d\sigma_{th}}{d\Omega} = \cos^{2}(\theta)\cos^{2}(\phi - \phi_{x}) + \sin^{2}(\phi - \phi_{x}),
\end{equation}    
where $\phi_{x}$ is the angle between the x axis and the polarization vector of the incoming x-ray photon, kept at zero in this work.

In IAM, we can study the scattering response as a function of pulse parameter by first calculating $R_j(t)$, $r_j(t)$ and $C_j(t)$.  These quantities can be readily computed using our Monte-Carlo/Molecular-Dynamics method, which has been detailed in previous work \cite{Ho-2017-JPB,Ho-2016-PRA}.   In brief, the interaction of the cluster of atoms with the incident XFEL pulse is treated quantum mechanically with a Monte Carlo method by tracking explicitly the time-dependent quantum transition probability between different electronic configurations. The total transition rate, $\Gamma$, between different electronic configurations $I$ and $J$ is given by
\begin{equation}
\Gamma_{I,J} = \Gamma^{P}_{I,J}+\Gamma^{A}_{I,J}+\Gamma^{F}_{I,J}+\Gamma^{RE}_{I,J}+\Gamma^{EI}_{I,J}+\Gamma^{RC}_{I,J}.
\end{equation}
Starting from the ground state of the neutral atom, we include the contribution from photoionization $\Gamma^{P}_{I,J}$, Auger decay $\Gamma^{A}_{I,J}$, fluorescence $\Gamma^{F}_{I,J}$, resonant excitation $\Gamma^{RE}_{I,J}$, electron-impact ionization $\Gamma^{EI}_{I,J}$ and electron-ion recombination $\Gamma^{RC}_{I,J}$.  Additionally, a molecular dynamics (MD) algorithm is used to propagate all particle trajectories (atoms/ions/electrons) forward in time.  We point out that we account for the non-dipole angular distribution of the photoelectrons, which is given by \begin{equation}
  \label{eq:angular}
\frac{\sin(\theta_e)^2}{(1 - (2*T_e/E_{rest})^{1/2}\cos(\theta_e))^4},
\end{equation}
where $T_e$ and $E_{rest}$ are the kinetic energy and rest mass of the photoelectron and $\theta_e$ is the emission angle with respect to the propagation axis of the incoming x-ray beam. For our small molecule and considered photon energies, this non-dipole emission does not has a large effect on the electron-impact ionization process.

By tracking the time evolution of electronic configurations of atoms and ions and their interaction with electrons, we investigated the effect of intense x-ray field on the single molecule scattering response.  To track the changes in electronic configuration and energetic photoelectron, a small time step of 0.1 attoseconds was used in the MC/MD calculations.  Since the probability of photoabsorption and nonlinear processes are small in an isolated molecule, a large number of replicas is needed to obtain a converged scattering profile for both the elastic and inelastic scattering channels.  For this study, we used 100 to 1000 replicas for each pulse parameter such that the error of the total scattering cross section is less than 0.1 \%.

\begin{figure*}[ht]
\centering
  \includegraphics[width=6.5 in] {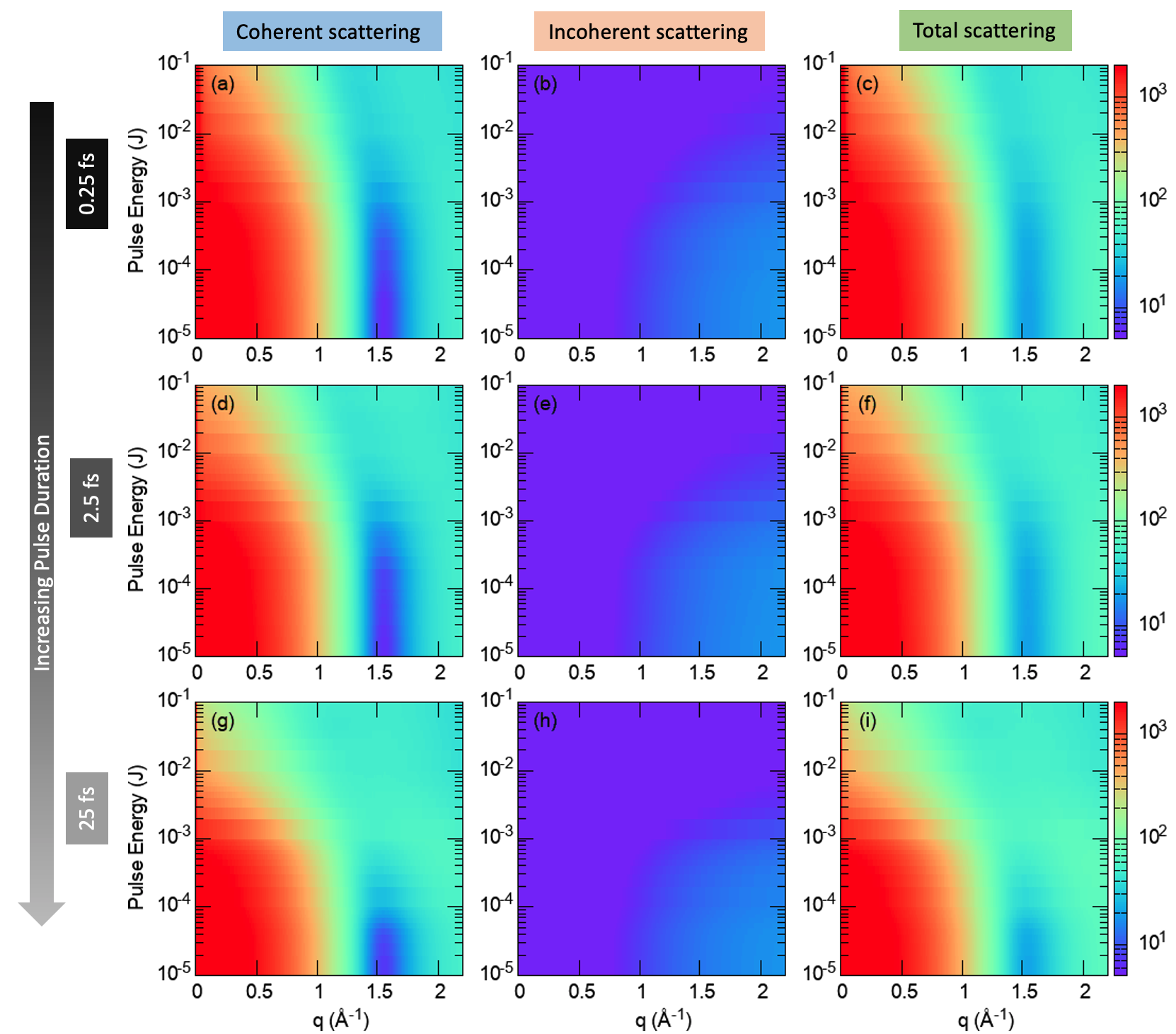}
  \caption{Pulse duration and pulse energy dependence of the azimuthally averaged differential x-ray scattering cross section of 1,3-cyclohexadiane exposed to an 5.6-keV pulse. Panel (a), (b) and (c) are the coherent, incoherent and the combined scattering cross sections calculated for a 0.25-fs pulse.  Panel (d), (e) and (f) are for a 2.5-fs pulse, whereas panel (g), (h) and (i) are for a 25-fs pulse.}
  \label{fgr:xproj_5600eV}
\end{figure*}

\subsection{Coherent Scattering Calculations Beyond the Independent Atom Model}\label{sec:adam}
In the high-field regime of coherent x-ray scattering, the non-linear x-ray interaction will include coherent scattering from core hole states generated by core ionizing photons within the pulse. The presence of a single core hole in a molecule can have a considerable effect on the surrounding electronic and nuclear environment, hence effecting the electron density undergoing the coherent scattering process. Therefore it is important to compare the coherent scattering of core hole states calculated by IAM with methods using bonding and correlation to relax the density in the presence of a core hole.

One step beyond IAM is the inclusion of molecular bonding by mean-field HF. There are two types of electron correlation missing from this method; static, arising from the interaction of near degenerate, electron configurations and dynamic, arising from the repulsive interactions in the many-electron environment. Errors attributed to both types of correlation can be addressed using methods going beyond a single determinant. Carrascosa \textit{et. al.} showed that multiconfigurational methods are necessary for accurate calculations of inelastic scattering, which depends on the electron correlation description in the reduced two-particle density matrix. However as the elastic scattering depends on the one electron density, it demands less from the electron correlation description and HF results deviated less from more computationally expensive multiconfigurational methods\cite{Carrascosa-2019-JCTC}. This makes HF a suitable method for exploring the coherent scattering beyond IAM for 1,3-cyclohexadiene, whereas multiconfigurational calculations, previuosly used on H2 and CO,  would become prohibitively expensive. 

DFT has gained wide popularity, as it addresses the electron correlation within a single determinant\cite{lee1988development} and maintains the same computational scaling as HF. All the complexities of the electron-electron interaction are placed in a single term, called the exchange-correlation functional. The exact form of which is unknown and is formed using approximate models. Shortcomings in DFT electron correlation can be attributed to the single determinant and delocalization error\cite{cohen2008insights}. The delocalization error in DFT manifests from the inability of the approximate model for the exchange-correlation energy to exactly cancel the spurious self-interacting Coulomb energy\cite{perdew1981self}. This results in the poor description of cation states with delocalized holes in both valence\cite{cheng2016charge} and core\cite{chong2007localized} regions.

We explore the effects of molecular bonding and electron correlation in the coherent scattering, by determining ground state and core-hole state wavefunctions by both Hartree-Fock and DFT. This enables the electronic and nuclear structures to relax in the presence of the core-hole. The triple-zeta 6-311+G$^{*}$ basis set was used throughout,  including both diffuse and polarization functions. A relatively small basis set can be justified as coherent elastic scattering is less dependent on the basis set size than inelastic scattering\cite{Carrascosa-2019-JCTC}, and vastly reduces the cost of integrating over the density, detailed below. Supplementary information (SI) Fig. S3 demonstrates that using the cc-pCVQZ basis set has little effect on the result.  All DFT calculations use the hybrid B3LYP exchange-correlation functional. Including a portion of Hartree-Fock exchange in the DFT exchange-correlation functional partially alleviates the delocalization error in the electron correlation\cite{cohen2008insights}. 

Initially the molecular geometry of the ground and core hole states were optimised in the QChem 5.3\cite{shao2015advances} software package using DFT with the core and valence orbitals treated separately by Boyes localisation\cite{foster1960canonical}. The maximum overlap method (MOM)\cite{besley2009self} was used to maintain the core hole state during the optimisation and three core-hole geometries were produced.

The optimized geometries were passed to a newly developed code for coherent elastic x-ray scattering calculations from the \textit{ab-initio} derived one electron density called X-RHO-SCATTER\cite{fouda2020xrho}. The code uses a modified version of the PSIXAS plugin\cite{ehlert2020psixas} using the Psi4NumPy development framework\cite{smith2018psi4numpy}, to generate DFT and Hartree-Fock wavefunctions for post-processing by ORBKIT\cite{hermann2016orbkit}. The core orbitals were treated with Pipek-Mezey localization\cite{pipek1989fast} and the core-hole was kept frozen during the optimization. ORBKIT generates the one electron density ($\rho_{1e}$(\textbf{r$_{1}$};\{R$_{A}$\})) on a grid from the wavefunction. This is required for the computation of coherent x-ray scattering as it facilitates numerical integration by curbature\cite{castro2015python} to generate the molecular form factor ($F_{mol}(\boldsymbol{q})$) by the following equation, 
\begin{equation}
  \label{eq:Fmol}
   F_{mol}(\boldsymbol{q}) = \int d^{3} \boldsymbol{r_{1}} \rho_{1e}(\boldsymbol{r}_{1};\{\textbf{R}_{A}\}) e^{i\boldsymbol{q} \cdot \boldsymbol{r}_{1}}.
\end{equation}




The differential cross sections are plotted in units of $r_{e}^2/radian^2$ (where $r_{e}$ is the classical electron radius). The maximum is in the forward direction and is proportional to the total number of electrons squared. Despite the relatively low cost of performing HF and DFT calculations on ground and core hole states, determining the differential coherent scattering cross section requires performing the integral in Eq.\ref{eq:Fmol}  N $\times$ M times, where N and M are the number of $\theta$ points $\phi$ points respectively. In order to observe significant changes in the coherent scattering with respect to the nuclear and electron structure, sufficient resolution between the scattering angles is required. In this study, scattering angles $\theta$ and $\phi$ are chosen as 100 points between 0 and 80 degrees for $\theta$ and 0 and 360 degrees for $\phi$.  To efficiently simulate the high resolution images, each integration was performed across a $18\times18\times18$ (\AA) grid of the density with a numerical precision of $10^{-3}$. It is shown in SI Fig. S4 that using higher precision integration ($10^{-4}$), yields the same physical result but at a higher contrast.


\begin{figure*}[ht]
\centering
  \includegraphics[width=6.5 in] {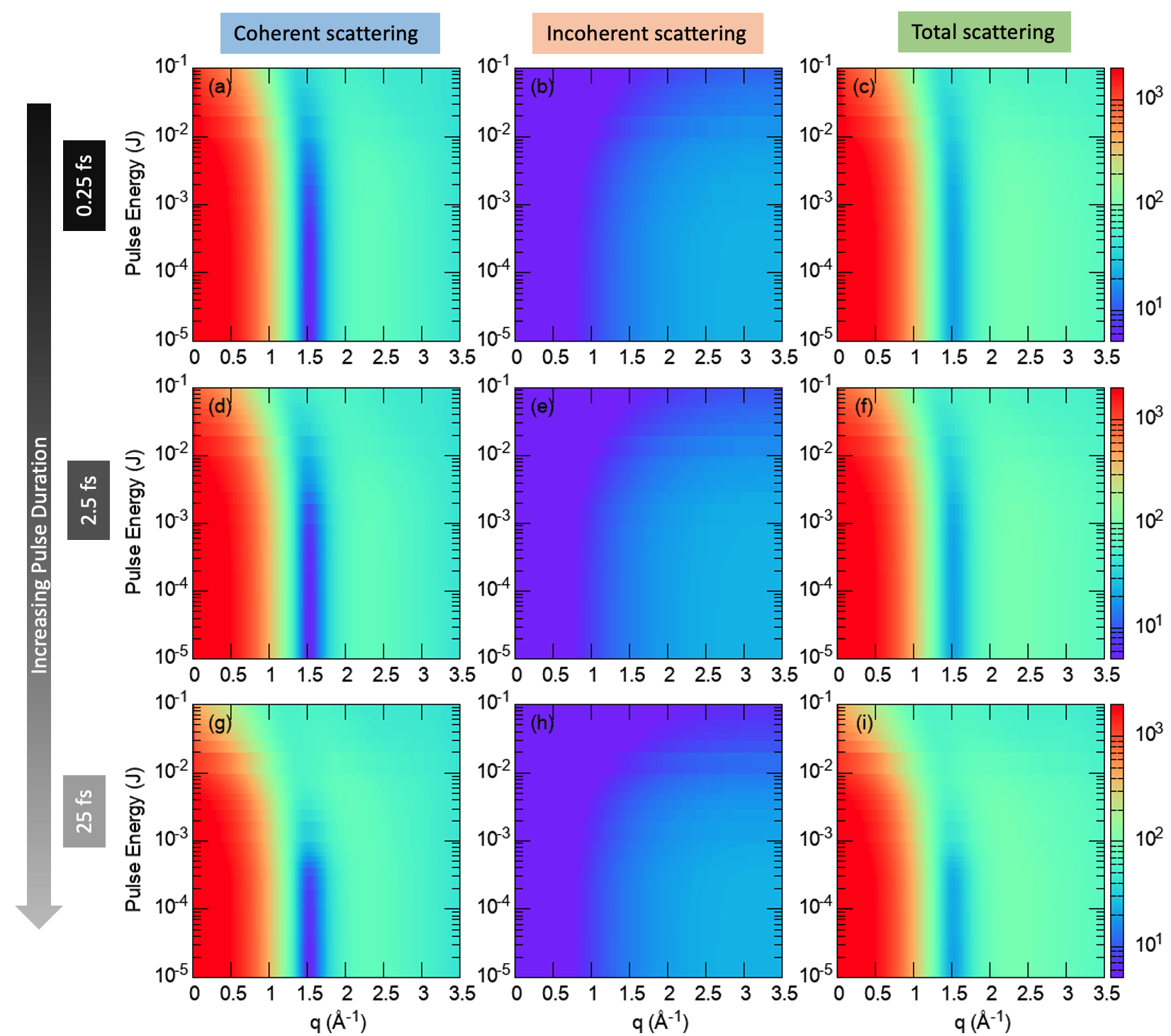}
  \caption{Pulse duration and pulse energy dependence of the azimuthally averaged differential x-ray scattering cross section of 1,3-cyclohexadiane exposed to an 9.0-keV pulse. Panel (a), (b) and (c) are the coherent, incoherent and the combined scattering cross sections calculated for a 0.25-fs pulse.  Panel (d), (e) and (f) are for a 2.5-fs pulse, whereas panel (g), (h) and (i) are for a 25-fs pulse.}
  \label{fgr:xproj_9000eV}
\end{figure*}

\begin{figure*}[!ht]
\centering
  \includegraphics[width=6.5 in] {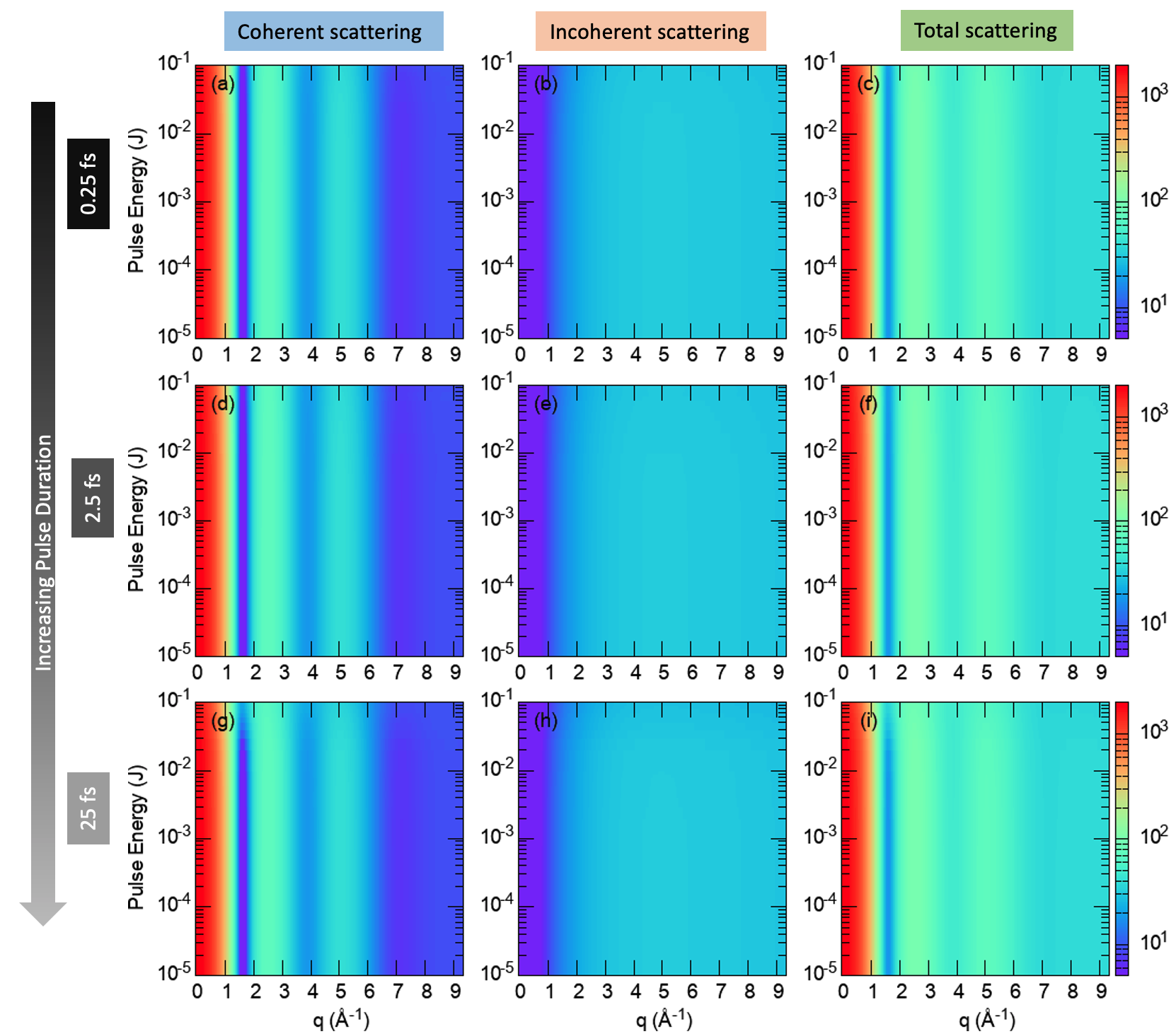}
  \caption{Pulse duration and pulse energy dependence of the azimuthally averaged differential x-ray scattering cross section of 1,3-cyclohexadiene exposed to an 24-keV pulse. Panel (a), (b) and (c) are the coherent, incoherent and the combined scattering cross sections calculated for a 0.25-fs pulse.  Panel (d), (e) and (f) are for a 2.5-fs pulse, whereas panel (g), (h) and (i) are for a 25-fs pulse.}
  \label{fgr:xproj_24000eV}
\end{figure*}

\begin{figure*}[!ht]
\centering
  \includegraphics[width=6 in] {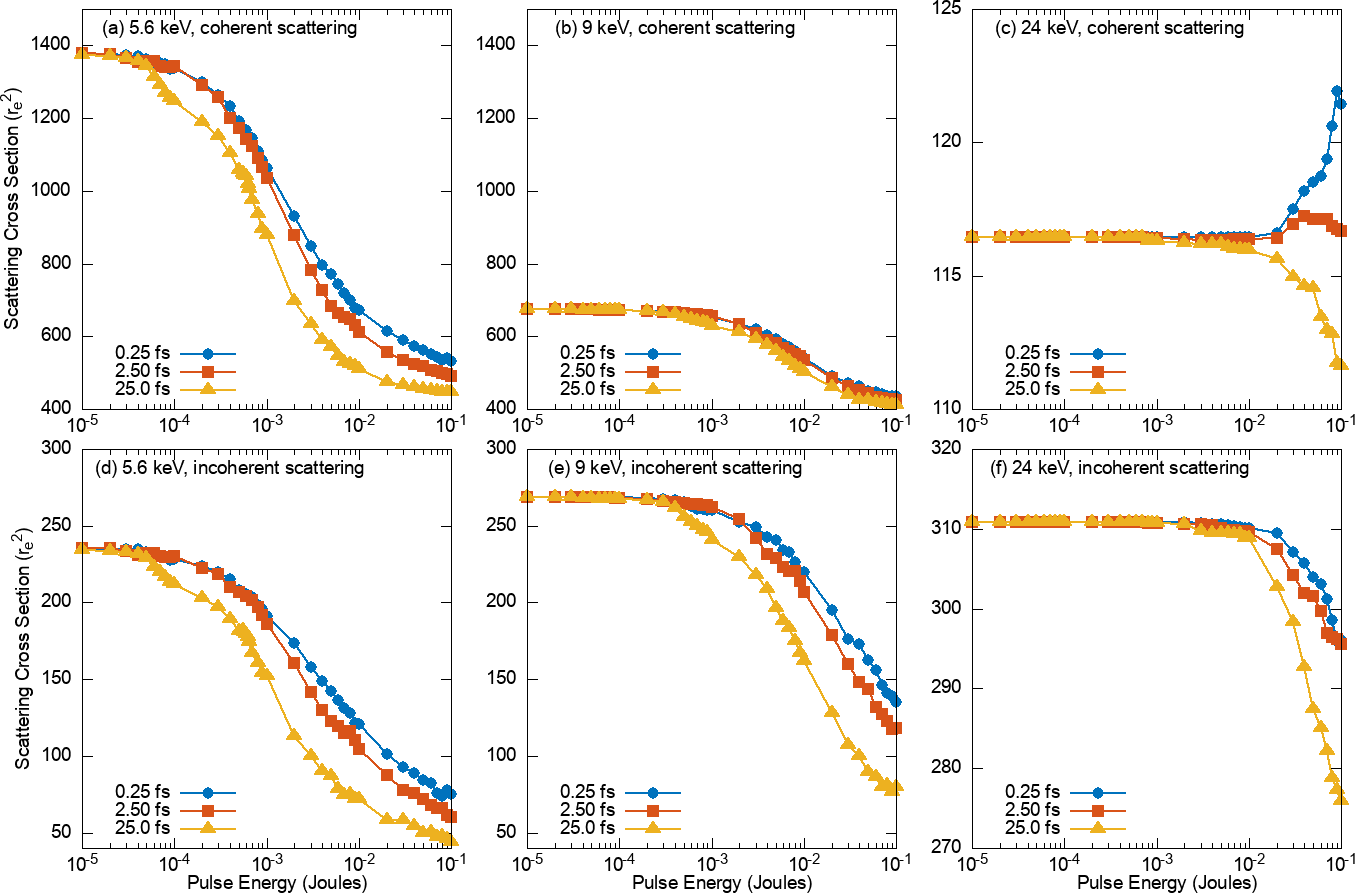}
  \caption{Photon energy, pulse duration and pulse energy dependence of elastic and inelastic scattering cross section.  Panel (a), (b) and (c) are the elastic scattering cross sections at 5.6, 9, and 24 keV respectively.  Panel (d), (e) and (f) are the inelastic cross section at 5.6, 9, and 24 keV respectively.  For both elastic and inelastic channel, the cross section scale (y-axis) at 24 keV is different than those at 5.6 and 9 keV to show their pulse energy dependence.}
  \label{fgr:TSCSall}
\end{figure*}

\begin{figure*}[!h]
\centering
  \includegraphics[width=7.4 in] {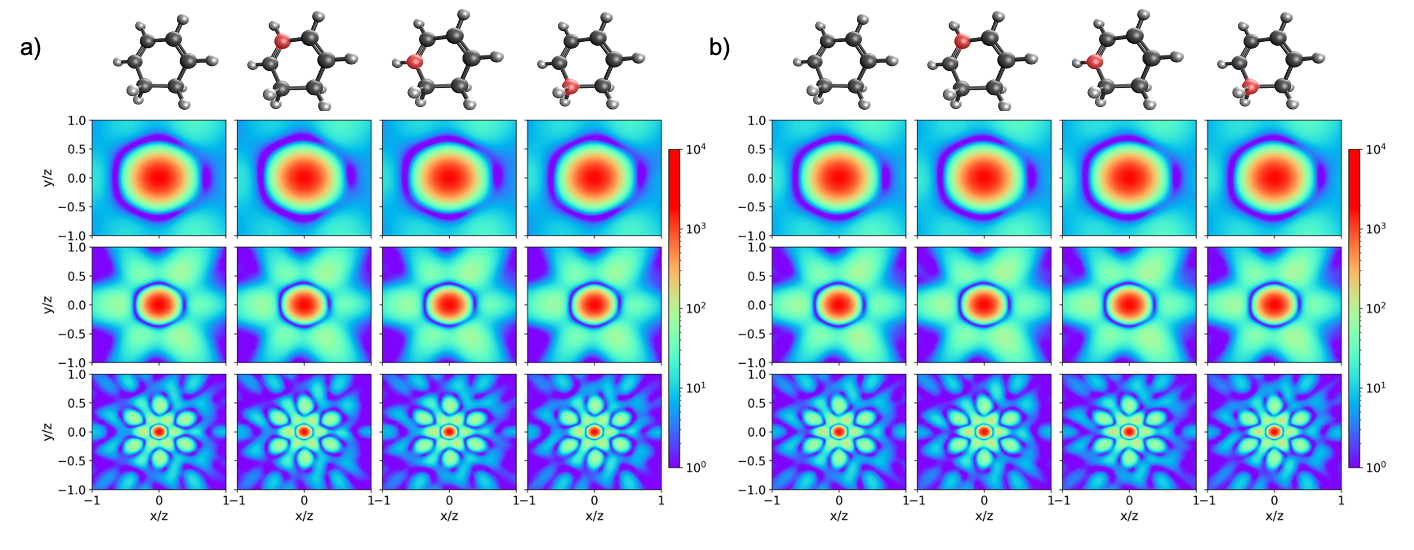}
  \caption{Detector images of the differential cross section for the coherent x-ray scattering of fixed-in-space 1-3-cyclohexadiene simulated by the IAM method in a) and by DFT (B3LYP 6-311+G$^{*}$) in b). The columns represent the electronic state and the rows the photon energy. The first column in both a) and b) is the neutral ground state and the next three are core ionised states, the carbon with the core hole is indicated in red at the top of each column. The first second and third rows show photon energies of 5.6, 9 and 24 keV respectively. The ground state optimised structure  was used throughout. }
  \label{fgr:AIMDFT}
\end{figure*}

\begin{figure*}[ht]
\centering
  \includegraphics[width=7.1 in] {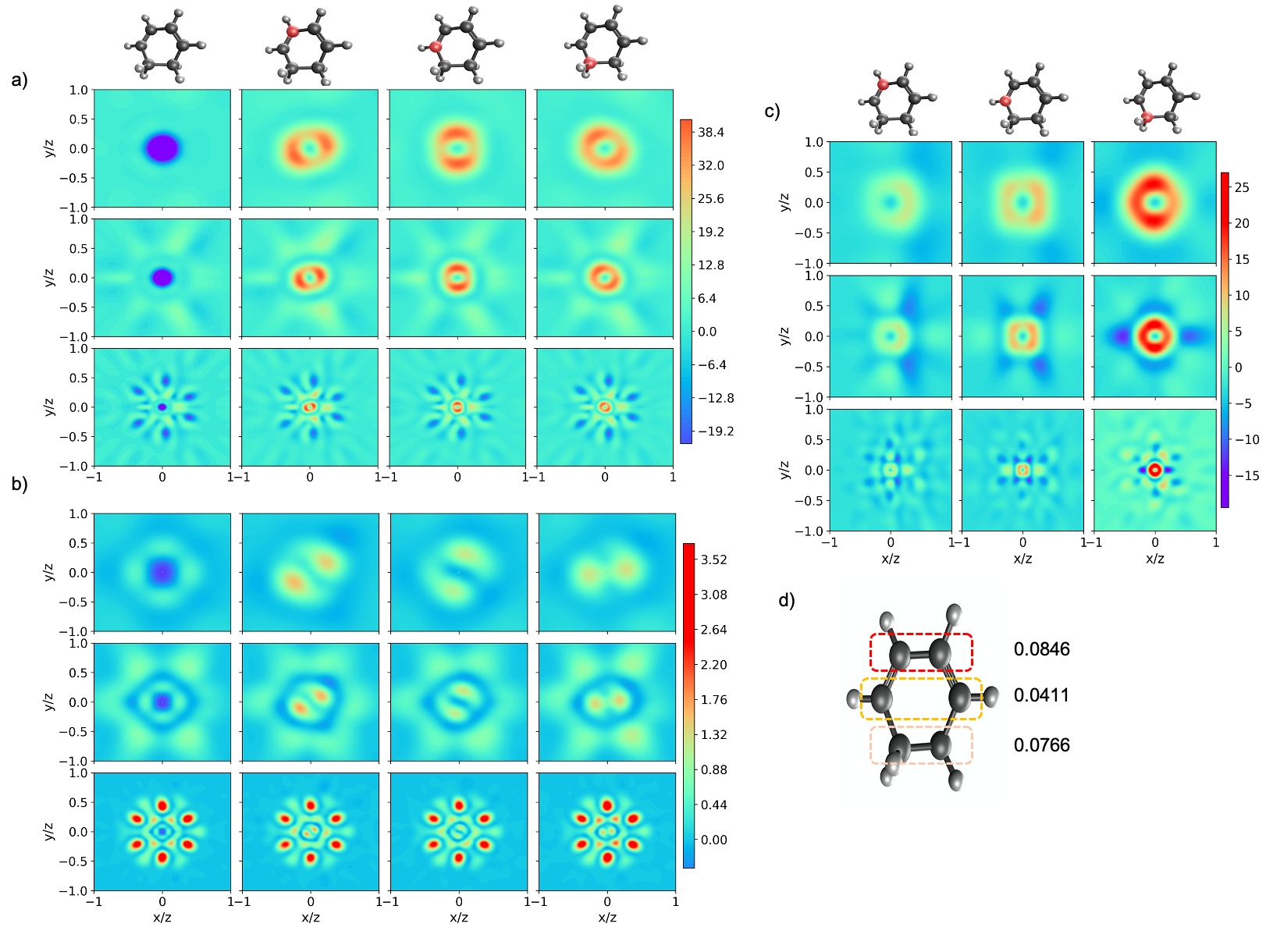}
  \caption{Comparisons of the coherent x-ray scattering calculations of 1-3-cyclohexadiene by difference detector images. The first second and third rows show photon energies of 5.6, 9 and 24 keV in a, b and c.  a) is the difference between the DFT (B3LYP 6-311+G$^{*}$) and IAM coherent scattering shown in Fig. \ref{fgr:AIMDFT}, b) shows the difference between the DFT core hole and HF (6-311+G$^{*}$) and c) the difference between the DFT core hole and ground stated optimized geometries.  The first column in both a) and b) is the neutral ground state and the next three, are core ionised states. The carbon with the core hole is indicated in red at the top of each column. The ground state geometry is used throughout a) and b). The columns in c) indicate the core hole present in the optimised geometry for each state.  d) Indicates the three carbon environments generating unique core hole structures from the DFT geometry optimisation used in c). The root mean squared deviation (RMSD) for core hole geometry, with respect to the ground state, is indicated to the right in Angstrom.}
  \label{fgr:AIMDFTHF}
\end{figure*}
\begin{figure*}[ht]
\centering
  \includegraphics[width=7 in] {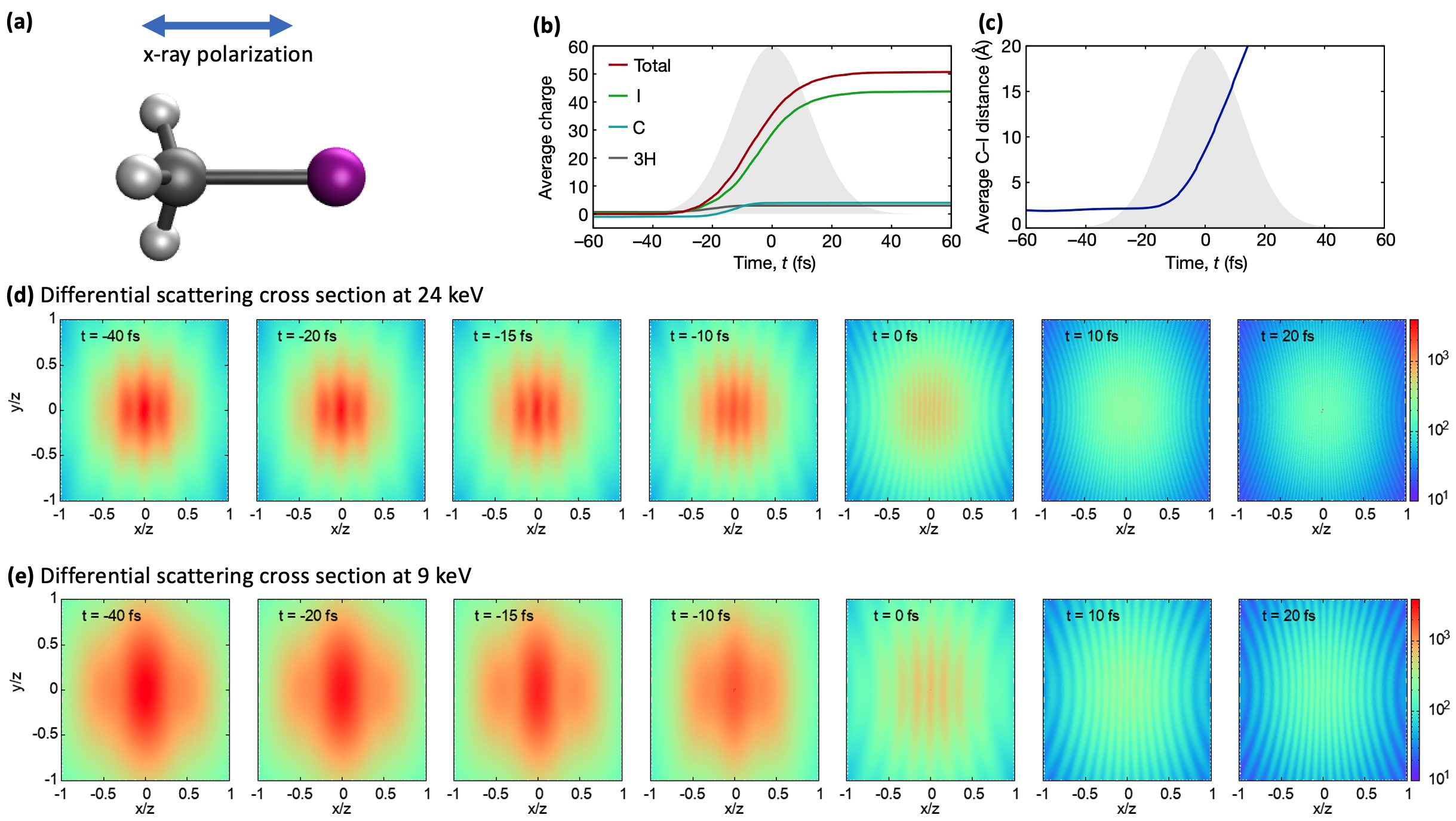}
  \caption{X-ray scattering images capturing x-ray excited molecular dynamics. (a) Scattering geometry of CH$_3$I.  (b) Charge transfer and (c) molecular dissociation dynamics of CH$_3$I induced by an intense 8.3-keV, 30-fs x-ray pulse with a fluence 5$\times$10$^{12}$ photons/$\mu$m$^2$ (reproduced from the work by Rudenko and coworkers \cite{Rudenko-2017-Nature}).  (d) X-ray scattering snapshots of time dependent dynamics of CH$_3$I in (b) and (c) captured by a 24-keV pulse.  (e) same as (d), but for a 9-keV pulse.   }
  \label{fgr:ch3i}
\end{figure*}

\section{Results and Discussion}\label{sec:Results}
\subsection{X-ray scattering from 1,3-cyclohexadiene: linear to nonlinear regime}\label{ssec:Results-MCMD}

Using the MC/MD method, we studied x-ray scattering at three x-ray energies:  5.6, 9.0 and 24 keV, motivated by the recently demonstrated single spike operation at 5.6 and 9.0 keV \cite{Marinelli-2017-APL,Huang-2012-PRL}, and the impending operation of LCLS-HE at higher x-ray energies.  We calculated for three pulse durations 0.25, 2.5 and 25 fs.  With an x-ray focal spot size of 100 nm and 45 pulse energies ranging from 10$\mu$J to 100 mJ, we have thus explored the scattering response for 1,3-cyclohexadiene from the linear to nonlinear absorption regime.  At these high x-ray energies, far above the K-edge of carbon, the saturation fluences of the molecule are $6.4\times10^{12} \mathrm{photons} /\mu\mathrm{m}^2$, $3.0\times10^{13}\mathrm{photons} /\mu\mathrm{m}^2$, and $7.0\times10^{14}\mathrm{photons} /\mu\mathrm{m}^2$ at 5.6, 9 and 24 keV respectively.  These saturation fluences correspond to pulse energies of 0.06 mJ, 0.44 mJ and 26.93 mJ per (100-nm)$^2$ and thus at all three photon energies the nonlinear absorption regime is reached.

Figures \ref{fgr:xproj_5600eV}, \ref{fgr:xproj_9000eV}, and \ref{fgr:xproj_24000eV} show scattering for photon energies of 5.6 keV, 9 keV and 24 keV, respectively. The columns represent the azimuthally averaged differential cross section $\frac{1}{2\pi} \langle \frac{d  \sigma_{total}}{d\Omega}(\boldsymbol{q}) \rangle_\phi$ as a function of q for coherent scattering (left), incoherent scattering (center) and the sum of these two channels (right).  The rows represent different pulse durations: 0.25 fs (top), 2.5 fs (middle) and 25 fs (bottom).  These results correspond to a fixed orientation of the molecule, in which the molecule ring lies on the x-y plane, and the incident x-ray propagates along the z-axis with its polarization along the x-axis.  The carbon atoms with two hydrogen atoms attached lie along the x-axis, on the lower half of the molecule, as pictured in Fig.  \ref{fgr:Scattering} (b). Calculations for other molecular orientations are not shown here, as the results reflect similar effects of non-linear x-ray scattering (see SI for the results corresponding to other molecular orientations). The lowest value on the y-axis corresponds to weak-field (linear) scattering.  The x-axis range plotted in Figs. \ref{fgr:xproj_5600eV}, \ref{fgr:xproj_9000eV}, and \ref{fgr:xproj_24000eV} reflects the $q$ associated with the scattering angle of $45\deg$ as shown later. The maximum $q$ available at a given photon energy, $q_{max}=2k\mathrm{sin}(\theta/2)$ where $k=2\pi/\lambda$ is $q_{max} =$ 5.7, 9.1 and 24.3 \AA$^{-1}$ for 5.6, 9.0 and 24 keV respectively.

Simple inspection shows that the deviations from the linear scattering are smaller at shorter pulse durations. For the 5.6 keV data with 25-fs pulse duration, Fig. \ref{fgr:xproj_5600eV}, deviations from the linear scattering begin around the saturation pulse energy of 0.06 mJ.  The onset of nonlinear behavior occurs at higher pulse energies for shorter pulses - consistent with the concept of ``diffract-before-destroy''.  Similar trends are seen for the 9.0 keV data, Fig. \ref{fgr:xproj_9000eV} where the saturation fluence corresponds to 0.44 mJ.  At 24 keV, the saturation fluence occurs near the maximum of the pulse energy calculated and thus nonlinear trends are not as obvious on the plots.

The atomic framework of the 1,3-cyclohexadiene molecule is reflected in the coherent scattering patterns and most readily observed in the scattering with 24 keV x rays.  Maxima in scattering are observed at $q = 2.5$\AA$^{-1}$ and $5$\AA$^{-1}$, corresponding to the atom-atom distances for C-C next-nearest, next-next-nearest neighbors (2.45, 2.51 and 2.85 \AA) and C-C nearest neighbors (1.40 and 1.50 \AA).  For the scattering at the lower photon energies, the q-range is limited and the C-C nearest neighbor distances are not accessible.  However, at the high photon energy, the inelastic scattering becomes much more apparent, such that when observing total scattering (Fig. \ref{fgr:xproj_24000eV} rightmost column) the second maximum is washed out.

It is of interest to compare the coherent and incoherent scattering channels for these conditions. The total coherent and incoherent scattering cross sections for 5.6, 9 and 24 keV are shown in Fig. \ref{fgr:TSCSall}.  The columns represent the 3 different photon energies and the top and bottom rows represent elastic and inelastic scattering, respectively.  As one increases the photon energy, the importance of the inelastic channel increases - the ratio of $\sigma_{elastic}/\sigma_{inelastic}$ goes from 5.8 at 5.6 keV to 0.37 at 24 keV for low intensity linear x-ray scattering.  However, a dramatic increase for $\sigma_{elastic}$ occurs for ultrashort pulses at high intensity and high x-ray energy, 24 keV. The origin of this increase is that at 24 keV $\sigma_{elastic}$ of carbon (0.97 barn) is comparable to that of a single electron (0.6 barn).  Thus, when the sample is damaged, i.e. ionized, the overall cross section will increase because of the presence of more delocalized electrons.  For the a 0.25-fs pulse, these electrons are likely to remain within the x-ray focus and contribute to the coherent scattering.  But, for a longer pulse duration (25 fs), these delocalized electrons can escape beyond the x-ray focus and their contribution is minimal.  Of course, the enhanced coherent scattering cross section due to the free electrons within the irradiation volume do not yield additional molecular structure information.  

For a given incident photon energy of $E_{k}$, the peak of the inelastic scattering energy spectrum $E_{k'}$ can be estimated from the Compton equation
\begin{equation}
    E_{k}/E_{k'} = 1+ (E_{k} \alpha^2)(1-\cos(\theta)), 
\end{equation}
where $\alpha$ is the fine structure constant.  This quantifies the energy of the outgoing photon versus the scattering angle, $\theta$.  For the maximum detection angle (see Figures \ref{fgr:AIMDFT}-\ref{fgr:AIMDFTHF}) considered here of $\theta$ = 45 degrees, the peaks of the energy spectrum for Compton scattering for incoming photon energies of 5.6, 9 and 24 keV are found to be 5.58, 8.95 and 23.67 keV, respectively.




\subsection{Coherent Scattering Beyond the Independent Atom Model}\label{sec:compare}

Multiphoton absorption in the non-linear interaction of x-ray scattering, will result in coherent scattering from core hole states generated by core ionizing photons within the pulse. Therefore, this section compares coherent scattering of core hole states calculated by IAM against HF and DFT; methods allowing the electron density to relax in the presence of the core hole. 

Fig \ref{fgr:AIMDFT} a) and b) show detector images of the coherent scattering calculated by IAM and DFT (B3LYP 6-311+G$^{*}$) respectively. Each row indicates a calculation performed at 5.6, 9 and 24 keV and each column represents an electronic state. The first column being the the neutral ground state and the next three are core ionised states (core hole indicated in red). The log-scale cross sections show that the inclusion of molecular bonding and electron correlation by DFT, has a small effect on the simulated coherent x-ray scattering. In addition to this, in both methods, there is little observable change between the electronic states; between the ground and core hole states and between the location of the core hole in the latter. 

The scattering patterns observed are associated with the fixed-in-space molecular structure shown above each column.  Scattering at the three photon energies allow us to zoom in on the molecular structure.  At lowest resolution, 5.6 keV (top row), the first minimum reflects the shape of the molecule.  At 9.0 keV, we see as maxima the largest atom-atom distances in the molecule, i.e the next-next-nearest neighbor carbon atoms opposite from each other in the ring, at $\pm x$ and at $y \pm 30\deg$.  At 24 keV, we are able to observe the  scattering islands associated with scattering from next-nearest neighbor carbon atoms at $\pm y$ and $x \pm 30\deg$ and intensities associated with nearest neighbor scattering. The scattering patterns calculated for using IAM and DFT methods are very similar. There is a left-right asymmetry which is attributed to the hydrogen atoms on the alkane carbons being out of plane in opposite directions.  The scattering patterns calculated by HF and DFT with a core hole optimised geometry, are shown in the SI, Fig. S1 and are also very similar.


These changes between IAM, HF and DFT descriptions of the molecular electron density are best visualized as difference patterns.
Fig. \ref{fgr:AIMDFTHF} a) shows the differences in the coherent cross sections calculated by DFT and IAM (DFT-IAM), from Fig. \ref{fgr:AIMDFT} and Fig. \ref{fgr:AIMDFTHF} b) shows the differences between DFT and HF calculations (DFT-HF). These plots have the same row and column structure with respect to the state and photon energy as Fig. \ref{fgr:AIMDFT} and they identify the differences between the methods. The maximum difference between the DFT and IAM methods is around 43$r_e^2$ which is roughly $2\%$ of the total coherent scattering cross section of the ground state. Decreases associated with the scattering from the carbon cores are clearly observable in the 24 keV scattering pattern.  The sensitivity to electron correlation is much smaller - the maximum difference between DFT and HF is approximately 3 which is around $0.2\%$ of the coherent scattering cross section of the ground state. The small percentages are expected due to coherent scattering only being a function of the one electron density and both HF and DFT being single determinant methods. 

However, despite the small magnitudes, the shape of the differences have an intriguing aesthetic. There is a clear sensitivity on the presence of a core hole and its location. Going from the ground state differences to the core-hole state differences there is a change in the shape about the center, which rotates with respect to the position of the core hole. This effect occurs for both the difference between DFT and IAM and DFT and HF, though the shape of features changes between the two. Here we have demonstrated how the difference in the coherent scattering presents a visualisation into the shift in the electron density with respect to the description of the electron correlation. However an in-depth interpretation into the relationship between the shape and correlation is difficult. Creation of the core hole generates Z+1 charge effect at the core hole site\cite{cavalleri2005half}, generating relaxation and dynamic electron correlation effects in the system\cite{aagren1993relaxation}. Both HF and DFT allow relaxation but only correlation is included in DFT. The difference between HF and IAM, shown in the SI Fig. S2 and gives the same result as Fig. \ref{fgr:AIMDFTHF} a). The effects of correlation can be attributed to Fig. \ref{fgr:AIMDFTHF} b) and the dynamic correlation effects are well isolated by our calculations, which use a localised core hole, kept frozen during the optimization. This prevents fractional occupation about the degenerate core holes and mitigates the effect of the delocalization error in DFT\cite{cohen2008insights,chong2007localized}. With further bench-marking, plotting the difference in coherent scattering may provide useful for evaluating electron correlation in \textit{ab-initio} methods, requiring evaluation of the density from high-level multiconfigurational approaches. 

Fig. \ref{fgr:AIMDFTHF} c) shows that the coherent scattering differences between the core-hole states calculated by DFT at the core-hole and ground state optimised geometries (core - ground). Also showing a small magnitude of the difference but striking features, sensitive to the location of the core hole. The root mean squared deviation of the atomic positions (RMSD) values are shown in Fig. \ref{fgr:AIMDFTHF} d) for the three carbon environments optimising to unique structures. The RMSD between the ground (\textit{g}) and core ionised (\textit{c}) state geometries containing \textit{N} atoms can be calculated by the following formula,
\begin{equation}
    RMSD(g,c) = \sqrt{\frac{1}{N}\sum_{i=1}^{N}((g_{ix} - c_{ix})^{2} + (g_{iy} - c_{iy})^{2} + (g_{iz} - c_{iz})^{2}) }.
\end{equation}
The RMSD values are small but give a complex effect on the coherent scattering difference. The structural change (RMSD) is largest when the core hole is localized on one of the alkene carbons. This corresponds to a change in the next-nearest-neighbor distance of roughly $2\%$. Fig. \ref{fgr:AIMDFTHF} c) shows that the largest magnitude of the difference occurs when the core hole is in the alkane environment. Though the magnitude of the difference is a small percentage of the total coherent scattering, it is conceivable that there may be significant effect from the core-hole state nuclear relaxation in high intensity x-ray scattering experiments where large number of core holes can be produced.

\section{Snapshots of x-ray multiphoton-induced dynamics } \label{sec:CH3I}

In a study of the femtosecond response of the CH$_3$I molecule (shown in Fig. \ref{fgr:ch3i}a) to ultra-intense hard x-ray radiation ($>10^{19}\mathrm{W/cm^2}$, 8.3 keV, 1.1 mJ), a fascinating ultrafast intramolecular charge-transfer process was observed to occur on the $\sim10$ fs timescale caused by multiphoton absorption \cite{Rudenko-2017-Nature}.  In the experiment, the yields and kinetic energies of the ionic fragments were measured.   An unexpectedly high total charge was observed in the molecule compared to irradiation of an individual atom with the same absorption cross-section. The intense radiation conditions yielded a total molecular charge of 54+ (I$^{47+}$ + C$^{4+}$ + 3H$^+$), compared to the isolated Xe atom (48+). Using the XMOLECULE toolkit\cite{Inhester-2015-SD}, the charge state distribution could be explained by a recurrent charge redistribution during multiphoton ionization of the iodine atom. Photoionization of the 2s and 2p shells, followed by  Auger cascade creates high charge on the iodine site.  The charge imbalance thus created drives electrons from the CH$_3$ to the iodine atom to refill the holes on sub-femtosecond timescales. 

We explore the feasibility of tracking the formation of the ``molecular black hole'' \cite{rudenko-black-hole} as it happens during the x-ray pulse, which requires an 'instantaneous' probe of the system, unlike the ionic fragments that result from complex decay cascades and nuclear dynamics occurring in part after the pulse. In principle, x-ray photoemission and Auger electron spectroscopies can provide the signature of all the intermediate ionic species, with even a sensitivity to nuclear dynamics \cite{travnikova2016hard}, but considering the complexity of the Auger spectra following a single ionization step these methods are not realistic for processes involving tens of electrons. In the spirit of this paper, we consider as an alternative the opportunities offered via x-ray scattering.  The simulated evolution of the molecular geometry and charge distribution, taken from Ref. \cite{Rudenko-2017-Nature}, is shown in Fig. \ref{fgr:ch3i}b,c respectively.  From this data we have calculated the x-ray scattering pattern for 24 keV and 9 keV photon energies at several instants during the 30-fs ionizing pulse (snapshots), as shown in Figs.\ref{fgr:ch3i}d,e respectively.  By the peak of the 30-fs FWHM pulse, the maximal charge state distribution is almost reached and the CH$_3$ - I bond distance has increased from its equilibrium value to $\sim10$\AA.  While the early stages of the process, the two-slit diffraction from the CH$_3$ and I entities, is best followed by the 24 keV probe at short distances, the latter stages are more easily seen with the 9 keV photons.  The total intensity fades as electrons are lost from the molecule. 

Measuring such snapshots requires a x-ray pump/x-ray probe setup, with the probe pulse significantly shorter than the 30-fs pump pulse in order to capture the evolution of the bond distance. Generating such pulses has already been demonstrated in the hard x-ray regime \cite{Huang-2017-PRL}, and it should be possible to use the large tunability capabilities of modern XFEL facilities to generate 2-color, 2-pulse configurations with an intense pump pulse and an ultrashort probe (sub-femtosecond)\cite{lutman2016fresh,Lutman-2018-PRL}. Using two different photon energies for the pump and probe is essential to be able to isolate the scattering signal of the probe that creates the snapshots. A simple and robust way consists in using a filter in front of the scattering detector, such that scattering from the pump is absorbed while signal from the probe is transmitted to reach the detector. This was successfully demonstrated previously \cite{Ferguson-2016-ScienceAdv}, even with a small energy separation between the two pulses, but the contrast can be improved by increasing the photon energy difference. This would be especially important considering that the pump pulse is very intense and contains many more photons than the probe pulse. Coming back to our simulations, one can consider two very different potential cases. In the first situation, the pump and probe pulses are close in photon energies and around 8 keV in our case, well suited to both produce the multiphoton ionization and probe the latter parts of the dynamics where nuclear distances increase and the number of scattering electrons has dropped. There, it would be advantageous to use a  probe with lower photon energy, and arranging the photon energies around the K-edge of a metal foil (e.g. nickel or copper). On the other hand, the probe energy can be set much higher than the pump, hereby probing finely the onset of the dynamics, perhaps by using the fact that the third harmonic is also produced when generating the fundamental photon energy. Using a relatively thick aluminum foil would here allow excellent contrast between the two scattering signals. While such pulse combinations are in principle possible through complex manipulations of the electron bunch shape and trajectories through a series of variable gap undulators, another promising option is coming to the horizon at LCLS-II. A new end-station (the Tender X-ray Instrument - TXI) will aim to combine on a single target the output of two independent XFEL undulators, with micrometer accuracy and femtosecond synchronization \cite{TXI-website}. These advances will allow optimizing each pulse separately, with the pump arm tuned to produce the highest x-ray intensity and the probe arm providing intense sub-femtosecond pulses necessary to capture the evolution of the molecules. In all cases, extracting the molecular properties as they evolve during the pulse would only be possible if the structure in the scattering patterns is not lost through rotational averaging. As mentioned above, laser alignment techniques would be required, such as demonstrated on iodomethane at a XFEL recently\cite{amini2018photodissociation}.

\section{Summary and Outlook}\label{sec:Outlook}
We have explored the use of x-ray scattering to monitor x-ray multiphoton induced dynamical phenomena.  By calculating the coherent and incoherent scattering response of the model system 1,3-cyclohexadiene to ultraintense x-ray pulses at 5.6, 9.0 and 24 keV for three pulse durations (0.25, 2.5 and 25 fs) we observe that the shortest pulse duration allows extraction of molecular structure at fluences larger than the nominal saturation fluence --- in accordance with the concept of ``diffract-before-destroy".  At the highest photon energy the inelastic channel is an bothersome background, but is also the most readily discriminated via energy shift.  We further explore the sensitivity of the coherent scattering channel to electron correlation effects in molecules, as has been previously discussed for atoms \cite{Jung-1998-PRL} and find that the observation of correlation effects require precise measurements on the order of 0.2\%.  Finally, we describe x-ray pump / x-ray probe methods that can directly probe ultrafast intramolecular charge transfer and dissociation induced by x-ray multiphoton absorption that were previously deduced from ion charge state measurements.

\section*{Conflicts of interest}
There are no conflicts to declare.

\section*{Acknowledgements}
This material is based on work supported by the U.S. Department of Energy, Office of Basic Energy Sciences, Division of Chemical Sciences, Geosciences, and Biosciences through Argonne National Laboratory. Argonne is a U.S. Department of Energy laboratory managed by UChicago Argonne, LLC, under contract DE-AC02-06CH11357.



\balance


\bibliography{main} 
\bibliographystyle{rsc} 

\end{document}


\maketitle

\begin{figure}[ht]
    \centering
    \centerline{\includegraphics[width=6.5 in]{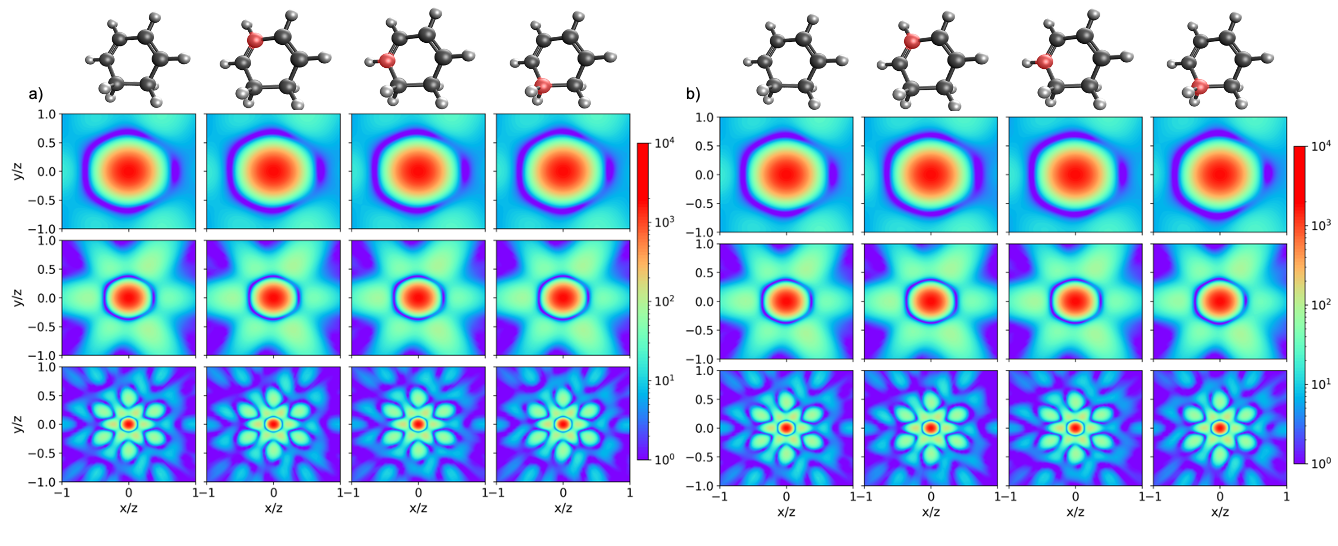}}
    \caption{Detector images of the differential cross section for the coherent x-ray scattering of fixed-in-space 1-3-cyclohexadiene simulated by the HF method in a) and by DFT (B3LYP 6-311+G$^{*}$), with optimised core hole states in b). The columns represent the electronic state and the rows the photon energy. The first column in both a) and b) is the neutral ground state and the next three are core ionised states, the carbon with the core hole is indicated in red at the top of each column. The first second and third rows show photon energies of 5.6, 9 and 24 keV respectively.}
    \label{HFDFTopt}
\end{figure}

\begin{figure}[ht]
    \centering
    \centerline{\includegraphics[width=6.5 in]{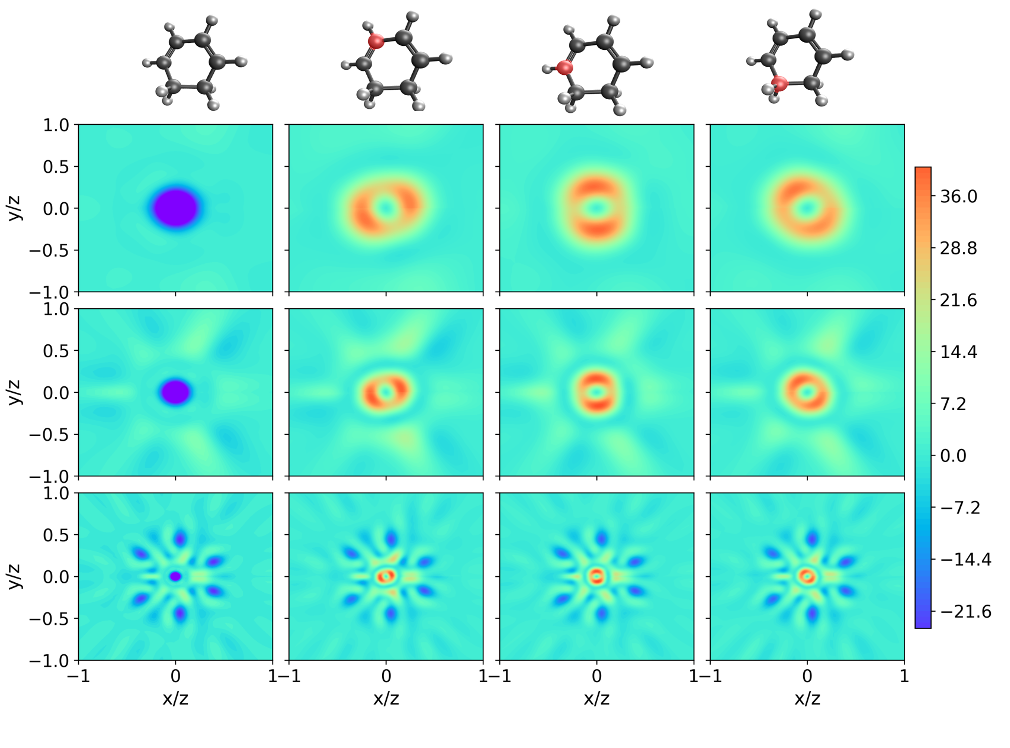}}
    \caption{Comparisons of the coherent x-ray scattering calculations of 1-3-cyclohexadiene by difference between the HF(6-311+G$^{*}$) and IAM. The first second and third rows show photon energies of 5.6, 9 and 24 keV respectively throughout. The first column in both a) and b) is the neutral ground state and the next three, are core ionised states. The carbon with the core hole is indicated in red at the top of each column. The ground state geometry is used throughout}
    \label{HFDFTopt}
\end{figure}

\begin{figure}[ht]
    \centering
    \centerline{\includegraphics[width=6.5 in]{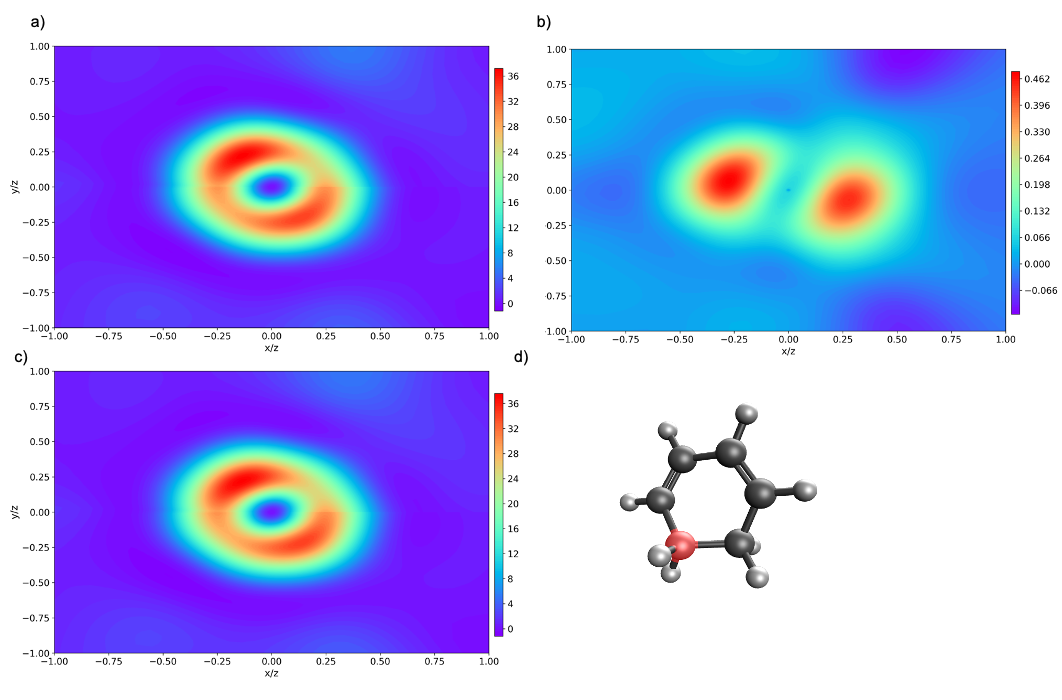}}
    \caption{a) Difference between HF(6-311+G$^{*}$) and IAM. b) The same as a) but using the cc-pCVQZ basis set in the HF calculation. c) Difference between HF(cc-pCVQZ) and HF(6-311+G$^{*}$). d) Core hole state used throughout calculations. All calculations at 5.6 keV.}
    \label{HFDFTopt}
\end{figure}

\begin{figure}[ht]
    \centering
    \centerline{\includegraphics[width=6.5 in]{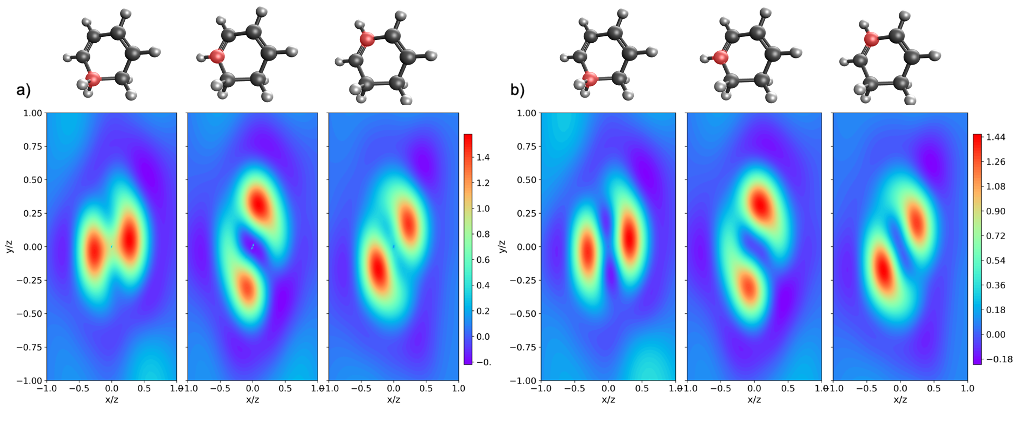}}
    \caption{Difference between the DFT(6-311+G$^{*}$) and HF(6-311+G$^{*}$) coherent scattering cross sections at three core hole states at 5.6 keV. a) Numerical integration precision of $10^{-3}$ (used throughout the main text, b Numerical integration precision of $10^{-4}$}
    \label{HFDFTopt}
\end{figure}

\begin{figure*}[ht]
\centering
  \includegraphics[width=6.5 in] {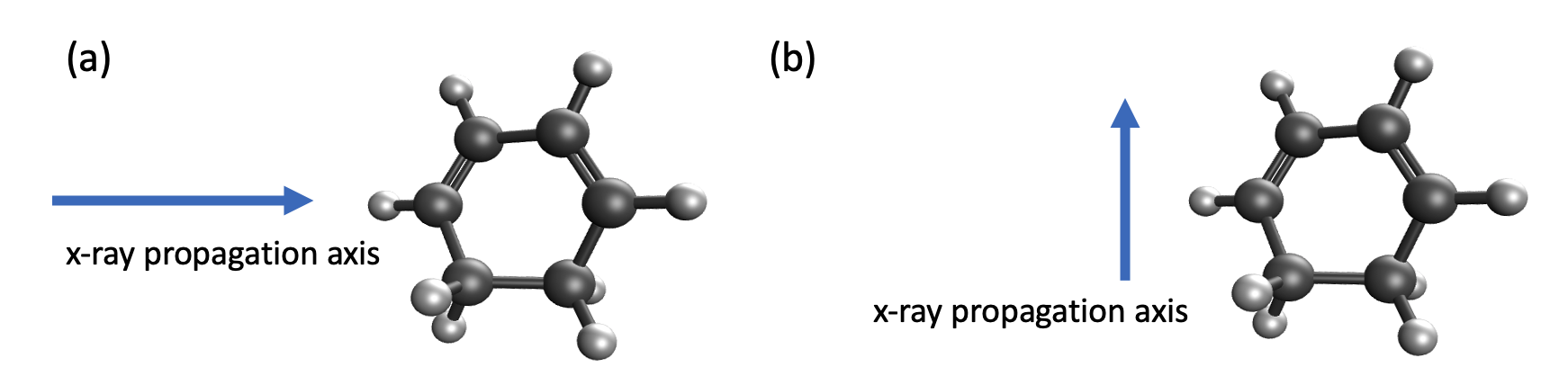}
  \caption{Two orientations of 1-3-cyclohexadiene with respect to the x-ray propagation axis.}
  \label{fgr:twoorientations}
\end{figure*}

\begin{figure*}[ht]
\centering
  \includegraphics[width=6.5 in] {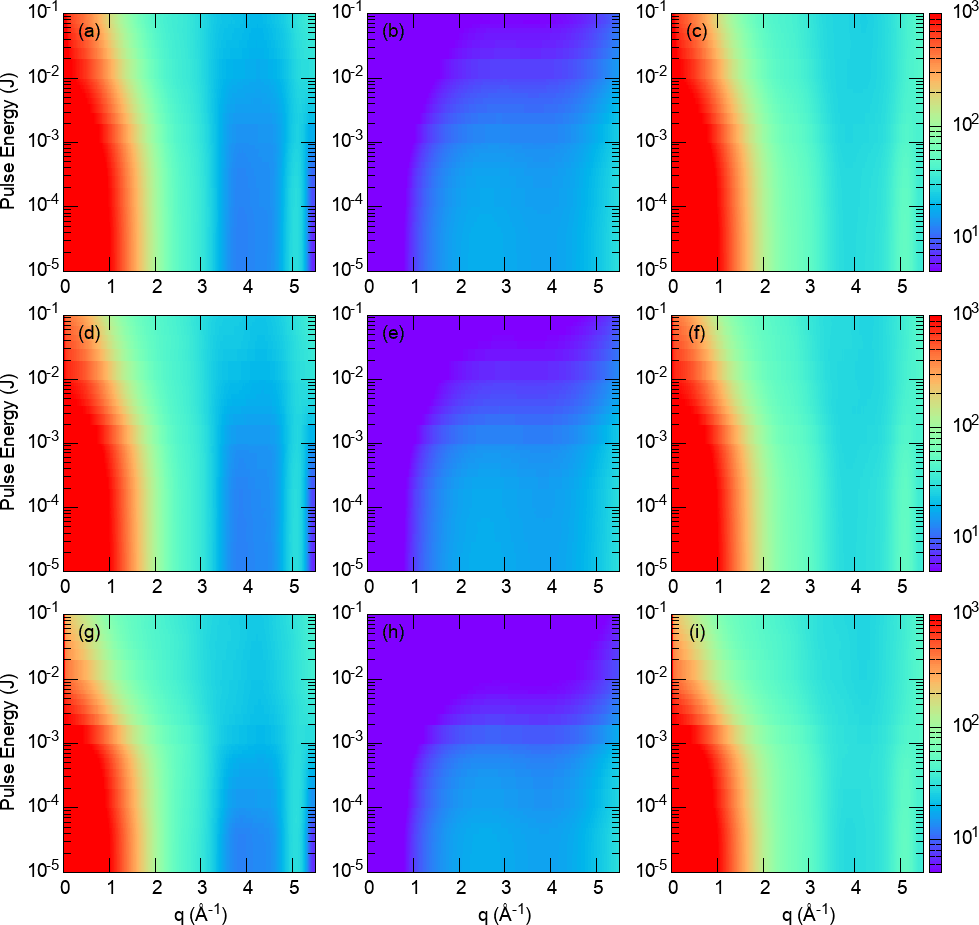}
  \caption{Pulse duration and pulse energy dependence of the azimuthally averaged differential x-ray scattering cross section of 1,3-cyclohexadiane exposed to an 5.6-keV pulse. Panel (a), (b) and (c) are the coherent, incoherent and the combined scattering cross sections calculated for a 0.25-fs pulse.  Panel (d), (e) and (f) are for a 2.5-fs pulse, whereas panel (g), (h) and (i) are for a 25-fs pulse.  The orientation of the molecule with respect to the x-ray is shown in Fig. \ref{fgr:twoorientations} (a).}
  \label{fgr:xproj_5600eV_Y}
\end{figure*}

\begin{figure*}[ht]
\centering
  \includegraphics[width=6.5 in] {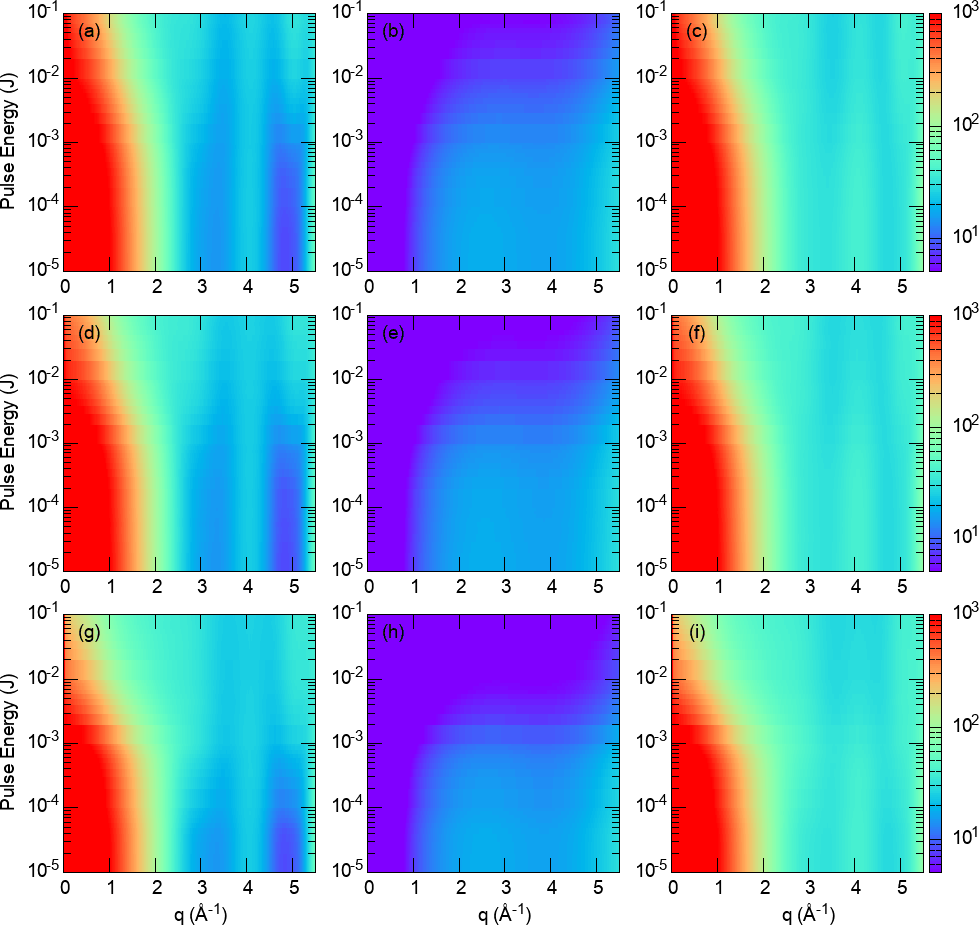}
  \caption{Same as Fig. \ref{fgr:xproj_5600eV_Y}, but the orientation of the molecule with respect to the x-ray is shown in Fig. \ref{fgr:twoorientations} (b).}
  \label{fgr:xproj_5600eV_Z}
\end{figure*}

\begin{figure*}[ht]
\centering
  \includegraphics[width=6.5 in] {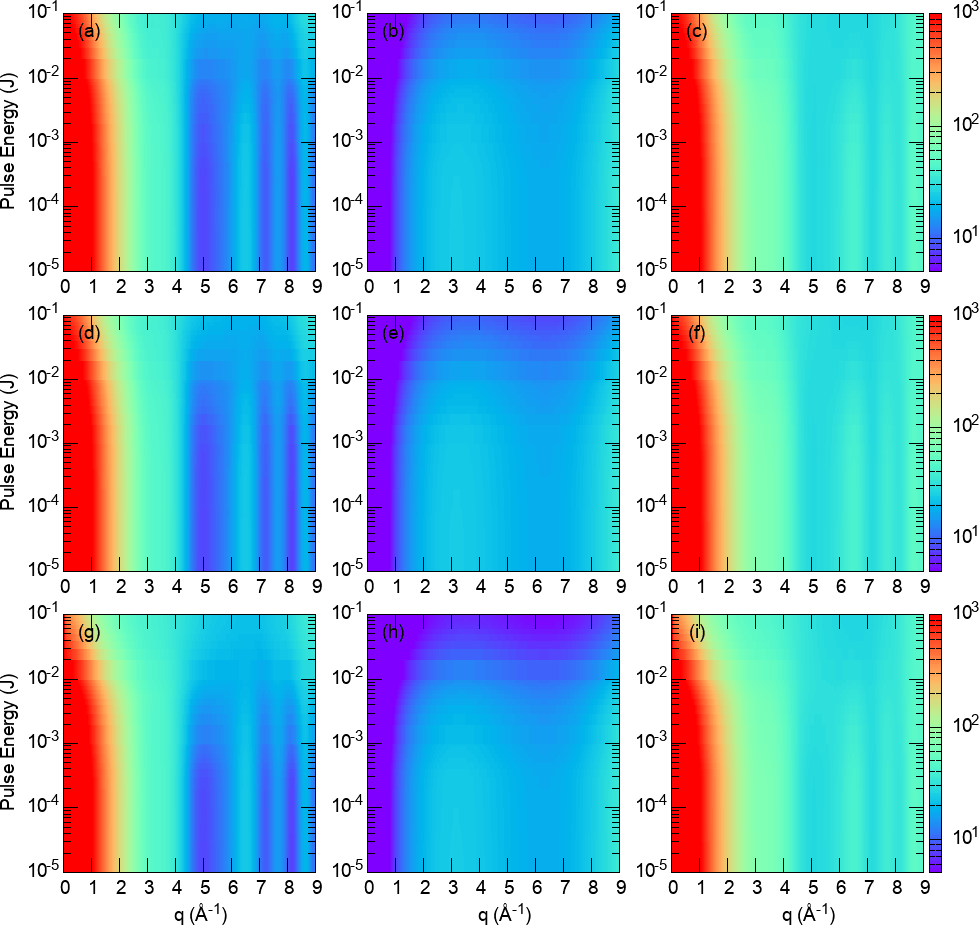}
  \caption{Same as Fig. \ref{fgr:xproj_5600eV_Y}, but the photon energy is 9 keV.}
  \label{fgr:xproj_9000eV_Y}
\end{figure*}

\begin{figure*}[ht]
\centering
  \includegraphics[width=6.5 in] {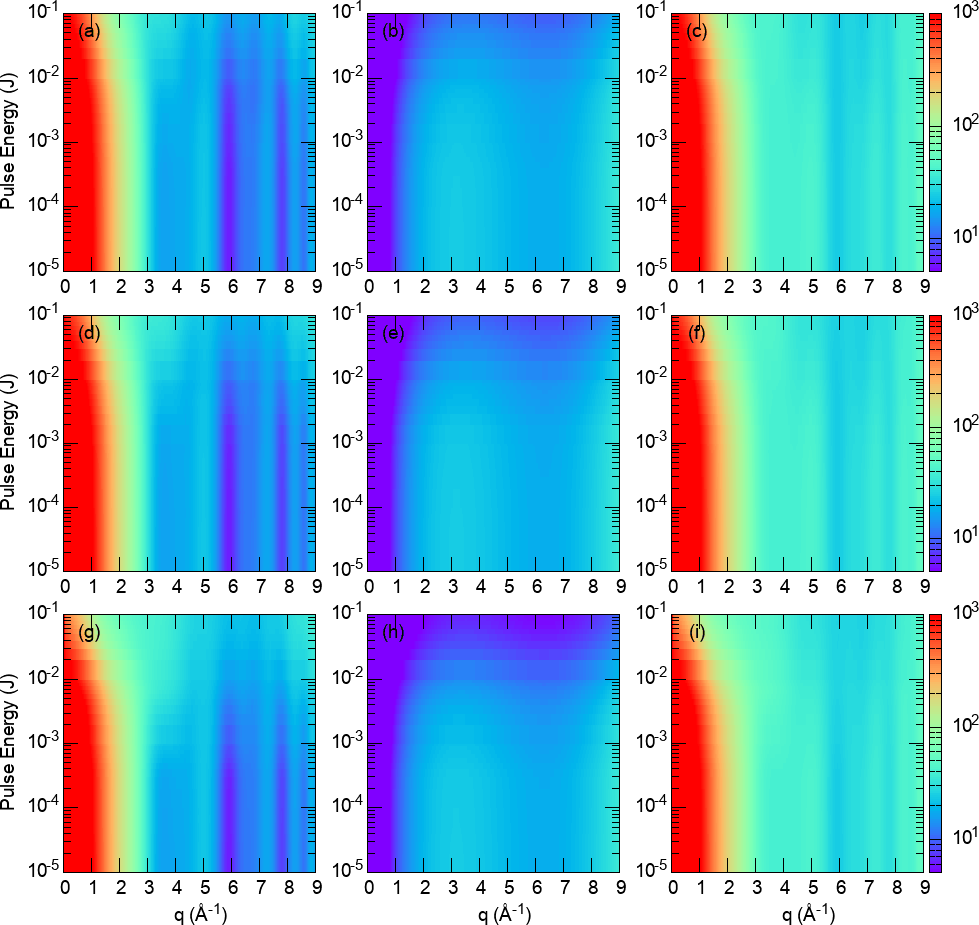}
  \caption{Same as Fig. \ref{fgr:xproj_5600eV_Y}, but the photon energy is 9 keV and the orientation of the molecule with respect to the x-ray is shown in Fig. \ref{fgr:twoorientations} (b).}
  \label{fgr:xproj_9000eV_Z}
\end{figure*}

\begin{figure*}[ht]
\centering
  \includegraphics[width=6.5 in] {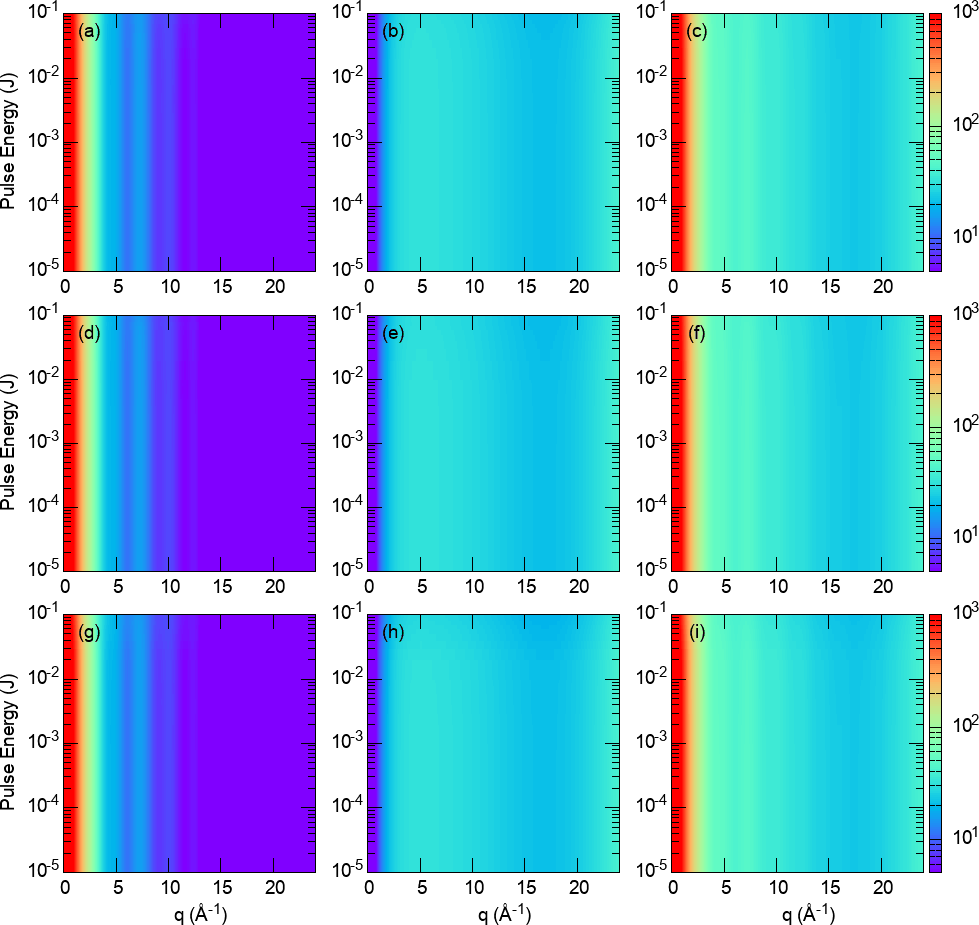}
  \caption{Same as Fig. \ref{fgr:xproj_5600eV_Y}, but the photon energy is 24 keV.}
  \label{fgr:xproj_24000eV_Y}
\end{figure*}

\begin{figure*}[ht]
\centering
  \includegraphics[width=6.5 in] {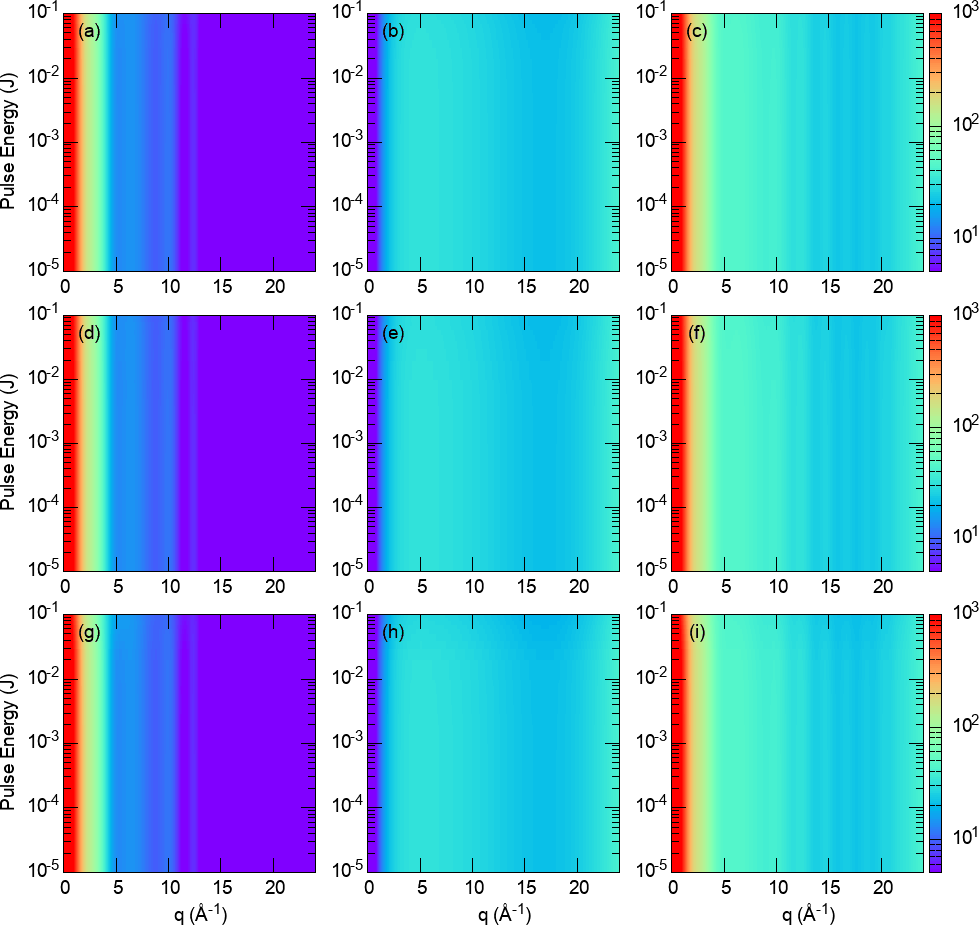}
  \caption{Same as Fig. \ref{fgr:xproj_5600eV_Y}, but the photon energy is 24 keV and the orientation of the molecule with respect to the x-ray is shown in Fig. \ref{fgr:twoorientations} (b).}
  \label{fgr:xproj_24000eV_Z}
\end{figure*}